\documentstyle[seceq,epsf,wrapft,preprint,ifthen]{ptptex}
\input{figure.mac}

\newcommand{\kyushuhet}[1]{
  \preprintnumber[3cm]{KYUSHU-HET-#1}
  }

\newcommand{\HIp}{H^{\prime}}
\newcommand{\Hp}{H^{\prime}}
\newcommand{\Heff}{H_{\rm eff}}

\newcommand{\JI}{J}
\newcommand{\Ub}[1]{\langle U(#1)\rangle}
\newcommand{\Ubi}[1]{\langle U(#1)\rangle^{-1}}
\newcommand{\cF}[1]{{\cal F}_{#1}}
\newcommand{\cG}[1]{{\cal G}_{#1}}

\newcommand{\cU}{{\cal U}}

\newcommand{\cV}[1]{{\cal V}_{#1}}
\newcommand{\cA}[1]{{\cal A}_{#1}}
\newcommand{\cM}[1]{{\cal M}_{#1}}
\newcommand{\cbM}[1]{\bar{{\cal M}}_{#1}}

\newcommand{\SF}[1]{S_{F}(#1)}
\newcommand{\SFI}[1]{S_{F}^{-1}(#1)}
\newcommand{\SFF}[2]{S_{F#1}(#2)}
\newcommand{\SFFI}[2]{S_{F#1}^{-1}(#2)}
\newcommand{\DF}[1]{\Delta_{F}(#1)}
\newcommand{\DFI}[1]{\Delta_{F}^{-1}(#1)}

\newcommand{\SZ}{Z_{3}}
\newcommand{\FZ}{Z_{2}}

\newcommand{\cO}{{\cal O}}
\newcommand{\ed}[1]{\frac{-i}{#1-i\epsilon}}
\newcommand{\inten}[1]{\int_{-\infty}^{\infty}\frac{d#1^{-}}{2\pi}}
\newcommand{\intenp}[1]{\int_{-\infty}^{\infty}\frac{d#1^{\prime -}}{2\pi}}
\newcommand{\HC}{\mbox{h.c.}}
\newcommand{\FTs}{\mbox{finite terms}}
\newcommand{\inp}[2]{#1\!\cdot\! #2}           

\newcommand{\slk}{\mbox{$k\hspace{-.5em}/$}}

\def\PRD#1{Phys.~Rev.~D~\andvol{#1}}

\title{
  A Construction of an Effective Hamiltonian from Feynman Diagrams :\\
  Application to the Light-Front Yukawa Model
  }
\author{
  Yuki YAMAMOTO\thanks{E-mail:~yuki1scp@mbox.nc.kyushu-u.ac.jp}
  }
\inst{
  Department of Physics, Kyushu University,
  Fukuoka 812-8581, JAPAN
  }
\abst{
  We study a similarity transformation to construct
  an effective Hamiltonian systematically,
  which does not contain particle-number-changing interactions,
  by means of Fukuda-Sawada-Taketani-Okubo's method.
  We show that such Hamiltonian can be constructed from Feynman diagrams
  and give rules for constructing it in the Light-Front Yukawa model.
  We prove that it is renormalized by the familiar covariant
  perturbative renormalization procedure.
  It is very advantageous that 
  the effective Hamiltonian can be
  obtained from our rules {\em immediately}.
  We also numerically 
  diagonalize it to the second order in the coupling constant
  as an exercise.
}

\begin{document}
\kyushuhet{45}

\maketitle

\section{Introduction}\label{sec:Introduction}
For many years, the bound state problem for QCD has been studied
very well but has not come to the complete understanding until today.
The main difficulty is that
the low energy dynamics of QCD needs nonperturbative calculations.
One of the convenient tools to solve the problem is a Hamiltonian approach in
Light-Front (LF) field theory which is quantized on the equal LF time surface.

An important feature of the LF field theory
is that the vacuum is trivial;
the Fock vacuum of the free part of the Hamiltonian is the true one.
It is very useful in the relativistic bound state problem
because we do not need to worry about solving the complex vacuum in contrast
to the usual equal-time (ET) field theory.
Even though the vacuum is trivial,
to solve the Schr\"odinger equation for the relativistic bound
state is difficult
because it is natural that the eigenstates are constructed
by the superposition of an infinite number of particle states allowed by
symmetries of the Hamiltonian.
Tamm-Dancoff (TD) approximation,\cite{T-D} \ 
which truncates the Fock space,
that is, limits the number of particles concerned with the interaction,
simplifies the practical calculations.
While TD approximation was originally proposed 
in the ET field theory,
Perry, Harindranath, and  Wilson suggested a LF version of it.\cite{P-H-W} \ 
They pointed out that since the truncated states do not consist a complete set
of the Hamiltonian, ultraviolet (UV) divergences are nonlocal and
noncovariant,
and counterterms which should cancel the divergences depend on the
sector of the Fock space within which they act.
This is called the problem of ``sector-dependent counterterms''.

A similarity transformation of the Hamiltonian does not change
the eigenvalues and is useful for getting the effective Hamiltonian.
There are two types of it.
One is the transformation in momentum space, which is equivalent to
integrating out states
which exchange energies more than some energy cutoff, proposed by
G{\l}azek and Wilson,\cite{G-W} \  and independently by Wegner.\cite{W} \ 
It is interesting that it gives the nonperturbative low energy physics
and a logarithmic confining potential in LF QCD,\cite{B-P} \ 
although it is hard to get the effective Hamiltonian
even in the lowest order in the coupling constant.
The details of this method and applications are discussed 
in Ref.~\citen{W-W-H-Z-P-G}
and recent progress is seen in Ref.~\citen{G-W_etc}.

Another is the transformation in the particle number space, which reduces 
the Hamiltonian to one which has no particle-number-changing interactions
so that the transformed one can be solved easily, and was proposed by 
Harada and Okazaki.\cite{H-O} \ 
It can avoid the problem of sector-dependent counterterms
because the origin of it is the general property of the Hamiltonian
that it has particle-number-changing interactions
and its eigenstates need an infinite number of particles.
But its actual calculations are complicated and tedious.
Although their method was considered in the LF field theory,
it is not new in the ET context
and has been used for getting the TMO potential
of nuclei.\cite{T-M-O} \ 
We call it Fukuda-Sawada-Taketani-Okubo's
(FSTO's) method.\cite{F-S-T,O} \ 
This method gives us an easier way for constructing the effective Hamiltonian
and seems to be promising in the LF framework.
However, it lacks manifest Lorentz covariance,
and therefore it is difficult to tell what sort of divergences the effective
Hamiltonian has before doing actual calculations.
It makes the renormalization procedure more complicated than 
the usual covariant perturbation theory.
It is highly desired to make transparent how the divergences emerge
in the FSTO's framework.

The purpose of this paper is to show that 
the effective Hamiltonian constructed by the FSTO's method
can be {\em immediately} obtained from Feynman diagrams
as the S-matrix element can
be in the covariant perturbation theory,
and that one can use the usual renormalization procedure
to renormalize it.
It makes the construction of the effective Hamiltonian systematic and easier.
Especially, it allows us to perform higher order calculations.\cite{Y} \ 

Our strategy is the following.
First, we show that the FSTO's effective Hamiltonian is a sum of 
the auxiliary operator $G$ and its products to the fourth order of the
interaction part of the Hamiltonian.
The advantages of using $G$ are that it is constructed by the same Feynman
diagrams as those for S-matrix elements, and that 
it has no particle-number-changing interaction so that one can easily
calculate normal-ordering of the products of those.
We give the rules for constructing $G$ from Feynman diagrams.
They are a little bit differ from the familiar Feynman rules 
in the covariant perturbation theory.
The set of those rules is one of the main results of the present paper.
Our method is more convenient and powerful
for constructing the effective Hamiltonian and 
discussing the renormalization of it than the other similarity methods.
Then, we find that there are three types of UV divergence.
One is the familiar loop divergence which
can be renormalized by the usual renormalization procedure.
The second comes from the difference between
our construction rules and Feynman ones.
We can make it harmless by using an ambiguity.
The above two only emerge in $G$.
The last emerges from the products of the renormalized $G$'s.
We show that it works as a box counterterm,\cite{G-H-P-S-W} \ 
which is needed to cancel the cutoff dependence of the eigenvalue,
in diagonalizing the effective Hamiltonian.

This paper is organized as follows.
In Sec.~\ref{sec:FSTO}, we briefly review the FSTO's method.
In Sec.~\ref{sec:RenEffH},
we show that the effective Hamiltonian constructed from the FSTO's method
is written in terms of the auxiliary operators $F$ or $G$.
$F$ is written in terms of $G$.
We consider the LF Yukawa model,\cite{G-H-P-S-W} \ in which Lagrangian
and Hamiltonian are given in Appendix~\ref{sec:LF_Yukawa},
as a concrete example,
and give the rules for the construction of $G$ from Feynman diagrams
and how to renormalize it and the effective Hamiltonian.
In Sec.~\ref{sec:Summary}, we summarize and discuss the validity of our method.
In Appendix~\ref{sec:NumericalResult}, as an exercise,
we calculate 
the eigenvalue of the ground states of the effective Hamiltonian up to
the second order in the coupling constant in the LF Yukawa model.
In Appendix~\ref{sec:Ambiguity}, we explain an ambiguity of the energy
integrations.

\section{Review of the FSTO's method}\label{sec:FSTO}
This section briefly reviews the FSTO's method partly following Ref.~\citen{O}.
The FSTO's method is to reduce a Hamiltonian to the block-diagonal form 
using a similarity transformation.
We can get the effective Hamiltonian for the subspace of the Fock
space without a loss of the necessary information.

We want to solve the Schr\"odinger equation 
\begin{equation}
  H|\Psi\rangle=E|\Psi\rangle,
\end{equation}
for the second-quantized Hamiltonian $H$
which consists of the free part $H_{0}$ and the interaction part $\Hp$:
\begin{equation}
  H=H_{0}+\Hp.
\end{equation}
The eigenstate $|\Psi\rangle$ can be expanded by the complete set of $H_{0}$.
To divide the Fock space into two, we introduce a projection operator $\eta$
which commutes with $H_{0}$:
\begin{equation}
  \left\{
    \begin{array}{rcl}
      \eta|\Psi\rangle&=&|\psi_{1}\rangle,\\
      (1-\eta)|\Psi\rangle&=&|\psi_{2}\rangle,
    \end{array}
  \right.
\end{equation}
where both $|\psi_{1}\rangle$ and $|\psi_{2}\rangle$ are
the states written in terms of the eigenstates of $H_{0}$.
$|\psi_{1}\rangle$ is the state in the target Fock space at our disposal.
For our purpose, we restrict $\eta$ to the one which selects the states
with definite number of particles.
In the matrix notation, $\eta$ and $|\Psi\rangle$ are written as
\begin{equation}
  \eta=\left(
    \begin{array}{cc}
      1 & 0\\
      0 & 0
    \end{array}
  \right),\quad
  |\Psi\rangle
  =\left(
    \begin{array}{c}
      |\psi_{1}\rangle\\
      |\psi_{2}\rangle
    \end{array}
  \right),
\end{equation}
respectively, and we express an arbitrary operator $O$ as
\begin{equation}
  O=\left(
  \begin{array}{cc}
    O_{11} & O_{12}\\
    O_{21} & O_{22}
  \end{array}
  \right),\quad
  \begin{array}{ll}
    O_{11}=\eta O\eta,&O_{12}=\eta O(1-\eta),\\
    O_{21}=(1-\eta)O\eta,&O_{22}=(1-\eta)O(1-\eta).
  \end{array}
\end{equation}

Let us consider the similarity transformation that leads the Hamiltonian
to the block-diagonal form in the following way
\begin{equation}
  \cU^{\dag}H\cU\
  \left(
    \begin{array}{l}
      |\psi_{1}\rangle^{\prime}\\
      |\psi_{2}\rangle^{\prime}
    \end{array}
  \right)
  =\left(
    \begin{array}{cc}
      \Heff & 0\\
      0 & *
    \end{array}
  \right)
  \left(
    \begin{array}{l}
      |\psi_{1}\rangle^{\prime}\\
      |\psi_{2}\rangle^{\prime}
    \end{array}
  \right),
  \label{eqn:SimilarityTr}
\end{equation}
where ``$*$'' abbreviates the part which is not needed for our purpose,
and
\begin{equation}
  \left(
    \begin{array}{l}
      |\psi_{1}\rangle^{\prime}\\
      |\psi_{2}\rangle^{\prime}
    \end{array}
  \right)
  =\cU^{\dag}
  \left(
    \begin{array}{l}
      |\psi_{1}\rangle\\
      |\psi_{2}\rangle
    \end{array}
  \right).
\end{equation}
$|\psi_{1}\rangle^{\prime}$ completely 
decouples from $|\psi_{2}\rangle^{\prime}$,
so that we can concentrate only on the equation for $\Heff$ and
$|\psi_{1}\rangle^{\prime}$ in the subspace selected by $\eta$:
\begin{equation}
  \Heff|\psi_{1}\rangle^{\prime}=E|\psi_{1}\rangle^{\prime}.
\end{equation}

We set an ansatz for the similarity (unitary) 
transformation operator
\begin{equation}
  \cU=\left(
  \begin{array}{cc}
    \cU_{11} & \cU_{12} \\
    \cU_{21} & \cU_{22}
  \end{array}
  \right)=
  \left(
  \begin{array}{cc}
    (1+A^{\dag}A)^{-1/2} & 
    -A^{\dag}(1+AA^{\dag})^{-1/2} \\
    A(1+A^{\dag}A)^{-1/2} &
    (1+AA^{\dag})^{-1/2}
  \end{array}
  \right).
  \label{eqn:ansatz}
\end{equation}
It is convenient to use $J$
\begin{equation}
  J=\left(
    \begin{array}{cc}
      1 & 0\\
      A & 1
    \end{array}
  \right),
\end{equation}
instead of $A$.
In order for the off-diagonal parts in Eq.~(\ref{eqn:SimilarityTr}) to
become zero, $J$ must satisfy
\begin{equation}
  (1-\eta)\left(\Hp J+[H_{0},J]-J\langle \Hp J \rangle\right)\eta=0,
  \label{eqn:EqOfJ}
\end{equation}
where we have introduced the notation $\langle\ \rangle$ as
\begin{equation}
  \langle O\rangle\equiv\eta O\eta+(1-\eta)O(1-\eta)
  =\left(
    \begin{array}{cc}
      O_{11} & 0 \\
      0 & O_{22}
    \end{array}
  \right).
\end{equation}
Using Eqs.~(\ref{eqn:SimilarityTr}), (\ref{eqn:ansatz}), and (\ref{eqn:EqOfJ}),
we formally obtain the effective Hamiltonian
\begin{equation}
  \Heff=\eta\langle J^{\dag}J\rangle^{-1/2}\langle J^{\dag}HJ\rangle
  \langle J^{\dag}J\rangle^{-1/2}\eta,
\end{equation}
but $J$ is unknown so far.

In the interaction picture, Eq.~(\ref{eqn:EqOfJ}) may be regarded as a 
differential equation
\begin{equation}
  (1-\eta)i\frac{d\JI(t)}{dt}\eta=(1-\eta)\left[\HIp(t)
    \JI(t)
    -\JI(t)\langle \HIp(t)\JI(t)\rangle\right]\eta,
  \label{eqn:DiffEqOfJ}
\end{equation}
where $\JI(0)$ is identified with $J$. 
$\Hp(t)$ and $J(t)$ are defined in this picture as
\begin{equation}
  \HIp(t)\equiv e^{iH_{0}t}\Hp e^{-iH_{0}t}e^{\epsilon t},\quad
  J(t)\equiv e^{iH_{0}t}J e^{-iH_{0}t}e^{\epsilon t},
\end{equation}
where  $e^{\epsilon t}$ is an adiabatic factor with $\epsilon\rightarrow 0+$.
We hereafter omit it in order to keep equations simple.
Provided that we set the initial condition
\begin{equation}
  \JI(-\infty)=1,
  \label{eqn:InitialCond}
\end{equation}
$J$ can be found order by order.
Alternatively, if we solve 
\begin{equation}
  i\frac{dV(t)}{dt}=\HIp(t)
  V(t)
  -V(t)\langle \HIp(t)V(t)\rangle,
  \label{eqn:EqOfV}
\end{equation}
under the initial condition
\begin{equation}
  V(-\infty)=1,
\end{equation}
$J$ is given by
\begin{equation}
  J=1+(1-\eta)V(0)\eta
  =\left(
    \begin{array}{cc}
      1 & 0\\
      V_{21}(0) & 1
    \end{array}
  \right).
  \label{eqn:J-V}
\end{equation}
It is easily found that the solution of Eq.~(\ref{eqn:EqOfV}) is given by
\begin{eqnarray}
  V(t)
  &=&U(t)\Ubi{t}\nonumber\\
  &=&\left(
    \begin{array}{cc}
      U_{11}(t) & U_{12}(t)\\
      U_{21}(t) & U_{22}(t)
    \end{array}
  \right)
  \left(
    \begin{array}{cc}
      U_{11}^{-1}(t) & 0\\
      0 & U_{22}^{-1}(t)
    \end{array}
  \right)
  \nonumber\\
  &=&\left(
    \begin{array}{cc}
      1 & U_{12}(t)U_{22}^{-1}(t)\\
      U_{21}(t)U_{11}^{-1}(t) & 1
    \end{array}
  \right),
  \label{eqn:SolutionOfV}
\end{eqnarray}
where $U(t)$ is the usual time evolution operator
\begin{equation}
  U(t)
  ={\rm T}\exp\left\{-i\int_{-\infty}^{t}dt^{\prime}\HIp(t^{\prime})\right\}.
\end{equation}
$U_{11}^{-1}(t)$ and $U_{22}^{-1}(t)$ are the inverse operators of
$U_{11}(t)$ and $U_{22}(t)$ respectively.
$J$ is explicitly
obtained from Eqs.~(\ref{eqn:J-V}) and (\ref{eqn:SolutionOfV}),
and then 
we expand it in the order of $\Hp$ to obtain the effective Hamiltonian
perturbatively:
\begin{equation}
  J=\left(
    \begin{array}{cc}
      1 & 0\\
      U_{21}(t)U_{11}^{-1}(t) & 1
    \end{array}
  \right)
  =1+\sum_{n=1}^{\infty}J_{n},
  \label{eqn:J}
\end{equation}
where $n$ is the order of $\Hp$.
The resultant effective Hamiltonian becomes
\begin{eqnarray}
  \Heff&=&\frac{1}{2}\eta\left(H_{0}+\Hp+
    \Hp J_{1}+\Hp J_{2}\frac{\mathstrut}{\mathstrut}\right.\nonumber\\
  &&\hspace{2em}\left.+\Hp J_{3}
    +\frac{1}{4}[J_{1}^{\dag}J_{1},\Hp J_{1}]\right)\eta+\HC+\cO(H^{\prime5}),
  \label{eqn:EffHamiltonian}
\end{eqnarray}
to the fourth order.
It is known that this can produce the TMO potential
in the symmetrical pseudoscalar pion theory with the pseudovector coupling,
which explains the properties of the deuteron very well.

Although Eq.~(\ref{eqn:EffHamiltonian}) 
is general and it is very straightforward to calculate it,
the calculation is absurdly tedious
because $J$ is not a time-ordered operator but a product of ones,
and has particle-number-changing interactions.
In order to renormalize the theory perturbatively,
we need to calculate the counterterms
but
it is not manifest what sort of the divergences emerge in 
the effective Hamiltonian before doing actual calculations.
It is not obvious either whether it is renormalizable or not
even in the renormalizable theory.

\section{Construction and renormalization of the effective Hamiltonian}\label{sec:RenEffH}
In this section, we give the construction method of the FSTO's effective
Hamiltonian from Feynman diagrams and discuss the renormalization of it.
First, we introduce the auxiliary operators $F$ and $G$ for convenience,
and then show that the effective Hamiltonian is written in terms of them.
$G$ can be constructed from the same Feynman diagrams as those
for S-matrix elements in the covariant perturbation theory.
We give the rules for the construction of it and
show that it is renormalized by the usual renormalization procedure.
Lastly, we show that the effective Hamiltonian has the divergent terms
even if $G$ is renormalized.
We discuss their role in diagonalizing the effective Hamiltonian.

\subsection{Definitions of $F$ and $G$}\label{sec:DefFG}
Let us define an operator
\begin{equation}
  F\equiv\eta\Hp J\eta,
  \label{eqn:F}
\end{equation}
motivated by the first few terms in Eq.~(\ref{eqn:EffHamiltonian}).
We will show that the effective Hamiltonian
can be rewritten in terms of $H_{0}$ and $F$.
By using Eqs.~(\ref{eqn:J-V})
and (\ref{eqn:SolutionOfV}),
we write $J$ as
\begin{equation}
  J=
  \left(
    \begin{array}{cc}
      0 & 0\\
      0 & 1
    \end{array}
  \right)
  +
  \left(
    \begin{array}{cc}
      1 & 0\\
      U_{21}(0)U_{11}^{-1}(0) & 0
    \end{array}
  \right)
  =(1-\eta)+U(0)\eta\Ubi{0}\eta.
  \label{eqn:J-U}
\end{equation}
Therefore, $F$ becomes
\begin{eqnarray}
   F&=&\eta\Hp U(0)\eta\Ubi{0}\eta\nonumber\\
   &=&G\eta\Ubi{0}\eta,
   \label{eqn:F-G-U}
\end{eqnarray}
where we define another operator $G$ as
\begin{eqnarray}
  G&\equiv&\eta\Hp U(0)\eta\nonumber\\
  &=&\eta{\rm T}\left\{\HIp(0)\exp\left[-i\int_{-\infty}^{0}dt\;
      \HIp(t)\right]\right\}\eta\nonumber\\
  &=&\eta\left\{\HIp(0)\frac{\mathstrut}{\mathstrut}\right.\nonumber\\
  &&\left.+\sum_{n=1}^{\infty}(-i)^{n}
    \int_{-\infty}^{0}dt_{1}\int_{-\infty}^{t_{1}}dt_{2}
    \cdots\int_{-\infty}^{t_{n-1}}dt_{n}\;
    \HIp(0)\HIp(t_{1})\HIp(t_{2})\cdots\HIp(t_{n})\right\}\eta.
  \label{eqn:G}
\end{eqnarray}
As we will shortly see, 
to introduce $G$ is crucial for using the the diagrammatic rules.
An advantage for using $G$ is that it is diagonal in the particle number
space so that we can normal-order the products of $G$'s more easily
than those of $J$'s and $\Hp$.
Note that $G$ is similar to the S-matrix operator
but there is one important difference:
the upper limit of the time integration is $0$.

We can find that $\eta\Ub{t}\eta$ is written in terms of $G$,
by integrating $t^{\prime}$ from $-\infty$ to $t$ after sandwiching $G$ 
between $e^{iH_{0}t^{\prime}}$ and $e^{-iH_{0}t^{\prime}}$:
\begin{eqnarray}
  &&-i\int_{-\infty}^{t}dt^{\prime}\;G(t^{\prime})\nonumber\\
  &=&-i\int_{-\infty}^{t}dt^{\prime}\;
  \eta\left\{\HIp(t^{\prime})\frac{\mathstrut}{\mathstrut}\right.\nonumber\\
  &&\left.+\sum_{n=1}^{\infty}(-i)^{n+1}
    \int_{-\infty}^{0}dt_{1}\int_{-\infty}^{t_{1}}dt_{2}
    \cdots\int_{-\infty}^{t_{n-1}}dt_{n}\;
    \HIp(t^{\prime})\HIp(t_{1}+t^{\prime})\HIp(t_{2}+t^{\prime})\cdots
    \HIp(t_{n}+t^{\prime})\right\}\eta\nonumber\\
  &=&\eta\left\{-i\int_{-\infty}^{t}dt^{\prime}\;\HIp(t^{\prime})\right.
  \nonumber\\
  &&\hspace{1em}\left.+\sum_{n=1}^{\infty}(-i)^{n+1}
    \int_{-\infty}^{t}dt^{\prime}
    \int_{-\infty}^{t^{\prime}}dt_{1}^{\prime}
    \int_{-\infty}^{t_{1}^{\prime}}dt_{2}^{\prime}
    \cdots\int_{-\infty}^{t_{n-1}^{\prime}}dt_{n}^{\prime}\;
    \HIp(t^{\prime})\HIp(t_{1}^{\prime})\HIp(t_{2}^{\prime})
    \cdots\HIp(t_{n}^{\prime})\right\}
  \eta\nonumber\\
  &=&\eta\Ub{t}\eta-\eta,
  \label{eqn:G-U}
\end{eqnarray}
where
\begin{equation}
  G(t)=e^{iH_{0}t}Ge^{-iH_{0}t}.
\end{equation}
By substituting $\langle U(0)\rangle$ in Eq.~(\ref{eqn:G-U})
to Eq.~(\ref{eqn:F-G-U}),
we can write $F$ as
\begin{eqnarray}
  F&=&G\left[\eta-i\int_{-\infty}^{0}dt\;G(t)\right]^{-1}\nonumber\\
   &=&G\left\{\eta+\sum_{n=1}^{\infty}\left[i\int_{-\infty}^{0}dt\;G(t)\right]^{n}\right\}.
\end{eqnarray}
It is important that $F$ depends only on $G$.
Here, we expand $F$ and $G$ in the order of $\Hp$:
\begin{equation}
  F=\sum_{m=1}^{\infty}F_{m},\quad G=\sum_{m=1}^{\infty}G_{m}.
\end{equation}
It is convenient to define
\begin{equation}
  \cF{n}=\sum_{m=1}^{n}F_{m},\quad \cG{n}=\sum_{m=1}^{n}G_{m}.
  \label{eqn:SumFG}
\end{equation}

Also, we solve the differential equation
 \begin{eqnarray}
  i\frac{d}{dt}(J_{1}(t)^{\dag}J_{1}(t))
  &=&-\eta\HIp(t)J_{1}(t)+J_{1}(t)^{\dag}\HIp(t)\eta\nonumber\\
  &=&-\eta\{\HIp(t)+\HIp(t)J_{1}(t)\}\eta
  +\eta\{\HIp(t)+J_{1}(t)^{\dag}\HIp(t)\}\eta\nonumber\\
  &=&-\cF{2}(t)+\cF{2}(t)^{\dag},
\end{eqnarray}
which is obtained from Eq.~(\ref{eqn:DiffEqOfJ}).
It is easily found that
\begin{equation}
  J_{1}^{\dag}J_{1}=i\int_{-\infty}^{0}dt\;\{\cF{2}(t)-\cF{2}^{\dag}(t)\},
  \label{eqn:J1J1}
\end{equation}
under the initial condition $J_{1}(-\infty)=0$.
From Eqs.~(\ref{eqn:F}), (\ref{eqn:SumFG}) and (\ref{eqn:J1J1}),
the term $\eta([J_{1}^{\dag}J_{1},H^{\prime}J_{1}]+\HC)\eta$ in $\Heff$
reads
\begin{eqnarray}
  &&[J_{1}^{\dag}J_{1},\eta\Hp J_{1}]+\HC\nonumber\\
  &=&[J_{1}^{\dag}J_{1},\eta\Hp\eta+\eta\Hp J_{1}]
  +[\eta\Hp\eta+J_{1}^{\dag}\Hp\eta ,J_{1}^{\dag}J_{1}]\nonumber\\
  &=&i\int_{-\infty}^{0}dt\;[\cF{2}(t)-\cF{2}^{\dag}(t),\cF{2}]+\HC
  \label{eqn:J-com}
\end{eqnarray}

Now we are ready to rewrite the effective Hamiltonian in terms of $H_{0}$
and $F$.
We finally obtain the effective Hamiltonian in the useful form
from Eqs.~(\ref{eqn:EffHamiltonian}), (\ref{eqn:F}), (\ref{eqn:SumFG})
and (\ref{eqn:J-com}):
\begin{equation}
  \Heff=\frac{1}{2}\eta\left\{H_{0}+\cF{4}+\frac{i}{4}\int_{-\infty}^{0}dt
    \;[\cF{2}(t)-\cF{2}^{\dag}(t),\cF{2}]\right\}\eta+\HC+\cO(H^{\prime5}).
  \label{eqn:FinalEffHamiltonian}
\end{equation}
All interactions are written in terms of
$\cG{n}$ through $\cF{n}$ to the fourth order in $H^{\prime}$.
To summarize, the effective Hamiltonian
can be easily constructed once $\cG{n}$ is obtained.

\subsection{Rules for the construction of $G$}\label{sec:RuleOfG}
In this subsection, we give the rules for the construction of $G$
from Feynman diagrams.
We do not need to use the old fashioned perturbation theory.
The knowledge in the covariant perturbation theory helps us to find them.

From Eq.~(\ref{eqn:G-U}), we immediately find that
\begin{equation}
 \eta S\eta=\eta U(\infty)\eta=
 \eta-i\int_{-\infty}^{\infty}dt\;G(t),
 \label{eqn:S-G}
\end{equation}
where $S$ is the familiar S-matrix operator.
Comparing order by order, one can relate the $n$th order term of $S$
to $G_{n}$ as
\begin{equation}
  \eta S_{n}\eta=-i\int_{-\infty}^{\infty}dt\;G_{n}(t).
  \label{eqn:cS-cG}
\end{equation}
It is important to make the difference between $\eta S_{n}\eta$ and $G_{n}$ clearer.
It is useful to rewrite them in terms of T-product and compare them
in order to find the correspondence with Feynman diagrams:
\begin{equation}
  \eta S_{n}\eta=\frac{(-i)^{n}}{n!}\int_{-\infty}^{\infty}dt_{1}
  \int_{-\infty}^{\infty}dt_{2}\cdots\int_{-\infty}^{\infty}dt_{n}\;
  \eta{\rm T}(\HIp(t_{1})\HIp(t_{2})\cdots\HIp(t_{n}))\eta,
\end{equation}
\begin{equation}
  G_{n}=\frac{(-i)^{n-1}}{(n-1)!}\int_{-\infty}^{0}dt_{1}
  \int_{-\infty}^{0}dt_{2}\cdots\int_{-\infty}^{0}dt_{n-1}\;
  \eta{\rm T}(\HIp(0)\HIp(t_{1})\HIp(t_{2})\cdots\HIp(t_{n-1}))\eta.
  \label{eqn:G_n}
\end{equation}
Wick's theorem tells us that T-products of $\HIp(t)$'s can be written
as sums of normal-ordered products of creation and annihilation operators
with the coefficients being amplitudes, which are given by
the same Feynman diagrams.
The difference between those arises from the time integrations.
It is apparent that the role of $\int_{-\infty}^{\infty}dt$ is to give 
the energy conserving delta function \mbox{$2\pi\delta(E_{out}-E_{in})$}
to each vertex in the Feynman diagrams, where $E_{out}$ is an outgoing
energy from the vertex and $E_{in}$ is an incoming energy to it. 
Because the time integrations end at $0$ in $G_{n}$,
the energy denominators \mbox{$(-i)(E_{out}-E_{in}-i\epsilon)^{-1}$}
appear instead of the delta functions.
The sign of an infinitesimal constant $\epsilon$ must be taken positive to 
ensure the convergence of the integrations at $t=-\infty$.
$n$ $\HIp(t)$'s become equivalent
and cancel the factor $1/n!$ in $S_{n}$ due to $n$ time integrations,
while there are $(n-1)$ time integrations and the factor $1/(n-1)!$
is canceled in $G_{n}$.
But there is a $\HIp(0)$ which is not integrated, therefore
there are $n$ terms for each vertex
which have $(n-1)$ products of the energy denominators.

Let us consider 
the LF Yukawa model%
\footnote{Lagrangian and Hamiltonian in this model
  are given in Appendix~\ref{sec:LF_Yukawa}.} 
as a concrete example.
In this model, the interaction part of the Hamiltonian 
has the form of
\begin{equation}
  \Hp=gH^{\prime(1)}+g^{2}H^{\prime(2)}.
\end{equation}
The effect of $H^{\prime(2)}$
which is the instantaneous interaction
is absorbed into the fermion propagator,
and we hereafter omit it
because we assume that 
LF diagrams are equivalent to the covariant one.%
\footnote{It is explained at the last paragraph 
  in Appendix~\ref{sec:LF_Yukawa}.}
We define $\eta$ as the projection operator for the two-body state
which consists of a fermion and an anti-fermion.
Because the power of
$g$ is equal to the number of the scalar fields
and \mbox{$\eta:(\mbox{odd}\;\phi$'s$):\eta=0$} is satisfied,
a product of $\Hp$'s satisfy
\begin{equation}
  \eta [\HIp(x_{1}^{+})\HIp(x_{2}^{+})\cdots\HIp(x_{n}^{+})]^{(m)}\eta=0,
\end{equation}
for odd $m$,
where the number in the brackets in the superscript means the order of $g$
and $x^{+}=(x^{0}+x^{3})/\sqrt{2}$ is LF time.
$\cF{2}$ and $\cF{4}$ become
\begin{equation}
  \cF{2}=\cG{2}+\cO(g^{4}),
  \label{eqn:cF2}
\end{equation}
\begin{equation}
  \cF{4}=\cG{4}+i\cG{2}\int_{-\infty}^{0}dx^{+}\cG{2}(x^{+})+\cO(g^{6}).
  \label{eqn:cF4}
\end{equation}
$G_{n}$ is written as
\begin{eqnarray}
  G_{n}&=&\sum_{\lambda}\int_{p_{1}}\int_{p_{2}}
  \cM{n}\;\eta b^{\dag}_{\lambda}(p_{1})b_{\lambda}(p_{2})\eta
  (2\pi)^{3}\delta^{3}(p_{1}-p_{2})\nonumber\\
  &&-\sum_{\lambda}\int_{p_{1}}\int_{p_{2}}
  \cbM{n}\;\eta d^{\dag}_{\lambda}(p_{1})d_{\lambda}(p_{2})\eta
  (2\pi)^{3}\delta^{3}(p_{1}-p_{2})\nonumber\\
  &&-\sum_{\lambda_{1}}\sum_{\lambda_{2}}\sum_{\sigma_{1}}\sum_{\sigma_{2}}
  \int_{p_{1}}\int_{p_{2}}\int_{l_{1}}\int_{l_{2}}\cV{n}\nonumber\\
  &&\hspace{5em}\times
  \eta b^{\dag}_{\lambda_{1}}(p_{1})d^{\dag}_{\sigma_{1}}(l_{1})
  d_{\sigma_{2}}(l_{2})b_{\lambda_{2}}(p_{2})\eta
  (2\pi)^{3}\delta^{3}(p_{1}+l_{1}-p_{2}-l_{2}),
  \label{eqn:cGn}
\end{eqnarray}
where  we define
\begin{eqnarray}
  \int_{p}&\equiv&\int
  \frac{dp^{+}d^{2}p_{\perp}}{2p^{+}(2\pi)^{3}}\theta(p^{+}),\\
  \delta^{3}(p-q)&\equiv&\delta(p^{+}-q^{+})\delta^{2}(p_{\perp}-q_{\perp}).
\end{eqnarray}
We hereafter call $p^{-}$ ``energy'' and
$(p^{+},p_{\perp})$ ``three-momentum''.
The creation and annihilation operators of the fermion,
$b^{\dag}_{\lambda}(p)$ and $b_{\lambda}(p)$, and ones of the anti-fermion,
$d^{\dag}_{\lambda}(p)$ and $d_{\lambda}(p)$, satisfy
the anti-commutation relations
\begin{equation}
  \{b_{\lambda_{1}}(p_{1}),b_{\lambda_{2}}^{\dag}(p_{2})\}
  =\{d_{\lambda_{1}}(p_{1}),d_{\lambda_{2}}^{\dag}(p_{2})\}
  =2p_{1}^{+}(2\pi)^{3}\delta^{3}(p_{1}-p_{2})\delta_{\lambda_{1}\lambda_{2}},
\end{equation}
and the other anti-commutators vanish.
The minus signs which are attached to
the second and third term of the right-hand
side of Eq.~(\ref{eqn:cGn}) come from the normal-ordered products
of the anti-fermion operators,
but are just the convention.
$\cM{n}$ and $\cbM{n}$ correspond to the amplitude for 
the fermionic and anti-fermionic one-particle state respectively,
and we have allowed for the conservation of the helicity.
As we discuss in Sec.~\ref{sec:DivergenceG},
these vanish under the physical (on-shell) renormalization condition.

$\cV{n}$ is a potential between a fermion and an anti-fermion in $G_{n}$
and is constructed by the $n$th order Feynman diagrams which have
an incoming fermion and anti-fermion and outgoing ones.
Since pair creations and annihilations of particles
from the vacuum are forbidden due to the merit of the LF field theory,
we must not include the disconnected diagrams.

There are three differences between our construction rules of $\cV{n}$ and
the usual Feynman rules for the general amplitudes:
\begin{itemize}
\item \label{enum:Vertex}
  Assign a factor
  \begin{equation}
    -g\gamma_{5}\frac{-i}{p_{out}^{-}-p_{in}^{-}-i\epsilon},
    \label{eqn:vertex}
  \end{equation}
  to each vertex, where $p_{out}^{\mu}$ ($p_{in}^{\mu}$) is
  the outgoing (incoming) four-momentum from (to) the vertex. 
  The energy denominator corresponds to the energy conserving delta function
  in the Feynman rules.

\item \label{enum:InternalLine}
  Assign an independent energy $p^{-}$ for each propagator and
  integrate it.
  We do not impose the conservation of the energies.

\item \label{enum:TotalEnergyFactor}
  Attach a total energy difference 
  $-(p_{1}^{-}+l_{1}^{-}-p_{2}^{-}-l_{2}^{-})$ 
  as an overall factor.
\end{itemize}  
It is obvious that the first two differences come from the domain of 
the time integrations.
The origin of the last one is that
there are $n$ terms which have different
products of the energy denominators as mentioned above.
For example, in the case of the diagram Fig.~\ref{fig:BoxDiagram}(a)
we consider the energy denominators obtained from Eq.~(\ref{eqn:G_n}) 
with factors of $i$:
\begin{eqnarray}
  &&(-i)^{3}i^{4}\left[
    \left(\frac{-i}{l_{1}^{-}+k_{2}^{-}-q_{1}^{-}-i\epsilon}\right)
    \left(\frac{-i}{q_{2}^{-}-k_{2}^{-}-l_{2}^{-}-i\epsilon}\right)
    \left(\frac{-i}{k_{1}^{-}-q_{2}^{-}-p_{2}^{-}-i\epsilon}\right)
  \right.\nonumber\\
  &&+
  \left(\frac{-i}{p_{1}^{-}+q_{1}^{-}-k_{1}^{-}-i\epsilon}\right)
  \left(\frac{-i}{q_{2}^{-}-k_{2}^{-}-l_{2}^{-}-i\epsilon}\right)
  \left(\frac{-i}{k_{1}^{-}-q_{2}^{-}-p_{2}^{-}-i\epsilon}\right)
  \nonumber\\
  &&+
  \left(\frac{-i}{p_{1}^{-}+q_{1}^{-}-k_{1}^{-}-i\epsilon}\right)
  \left(\frac{-i}{l_{1}^{-}+k_{2}^{-}-q_{1}^{-}-i\epsilon}\right)
  \left(\frac{-i}{k_{1}^{-}-q_{2}^{-}-p_{2}^{-}-i\epsilon}\right)
  \nonumber\\
  &&\left.+
    \left(\frac{-i}{p_{1}^{-}+q_{1}^{-}-k_{1}^{-}-i\epsilon}\right)
    \left(\frac{-i}{l_{1}^{-}+k_{2}^{-}-q_{1}^{-}-i\epsilon}\right)
    \left(\frac{-i}{q_{2}^{-}-k_{2}^{-}-l_{2}^{-}-i\epsilon}\right)
  \right].
  \label{eqn:ED1}
\end{eqnarray}
The factor $(-i)^{3}$ comes from 
three time integrations $(-i\int_{-\infty}^{0}dx^{+})$,
while the factor $i^{4}$ comes from four vertices
$(-ig\bar{\psi}\gamma_{5}\phi\psi)$.
In general, the total factor of $i$ is $(-i)^{V-1}i^{V}$ where $V$ is
the number of vertices in the diagram.
Since the sum is the combination of removing one from four denominators,
we can make it a product of all the energy denominators:
\begin{eqnarray}
  &=&-(p_{1}^{-}+l_{1}^{-}-p_{2}^{-}-l_{2}^{-})
  \left(\frac{-i}{p_{1}^{-}+q_{1}^{-}-k_{1}^{-}-i\epsilon}\right)
  \left(\frac{-i}{l_{1}^{-}+k_{2}^{-}-q_{1}^{-}-i\epsilon}\right)
  \nonumber\\
  &&\hspace{5em}\times
  \left(\frac{-i}{q_{2}^{-}-k_{2}^{-}-l_{2}^{-}-i\epsilon}\right)
  \left(\frac{-i}{k_{1}^{-}-q_{2}^{-}-p_{2}^{-}-i\epsilon}\right).
  \label{eqn:ED2}
\end{eqnarray}
This factor corresponds to the elimination of the total energy conserving
delta function.

Finally, $\cV{n}$ becomes
\begin{eqnarray}
  \cV{n}&=&-(p_{1}^{-}+l_{1}^{-}-p_{2}^{-}-l_{2}^{-})\sum_{\rm diagrams}
  \int_{-\infty}^{\infty}\frac{dk_{1}^{-}}{2\pi}
  \int_{-\infty}^{\infty}\frac{dk_{2}^{-}}{2\pi}\cdots
  \int_{-\infty}^{\infty}\frac{dq_{1}^{-}}{2\pi}
  \int_{-\infty}^{\infty}\frac{dq_{2}^{-}}{2\pi}\cdots
  \nonumber\\
  &&\times\left(\frac{-i}{p_{1}^{-}-k_{1}^{-}-q_{1}^{-}-i\epsilon}\right)
  \left(\frac{-i}{l_{1}^{-}-k_{2}^{-}+q_{2}^{-}-i\epsilon}\right)\cdots
  \nonumber\\
  &&\times(-i\cA{n}),
\end{eqnarray}
where $\cA{n}$ is the usual amplitude constructed from the Feynman diagrams
but each energy of the propagators is an independent variable
which is integrated outside of $\cA{n}$.
Therefore, $\cA{n}$ is constructed from the propagators $\SF{k}$ and $\DF{q}$
with each independent energy, and the vertex $-g\gamma_{5}$,
and the external fermion line $\bar{u}(p_{1},\lambda_{1})$, $u(p_{2},\lambda_{2})$,
$v(l_{1},\sigma_{1})$, $\bar{v}(l_{2},\sigma_{2})$,
and an integration of four-momentum for each loop.
The three-momenta are conserved but energies are not.

Although each propagator includes both the loop and independent energy,
it is easy to split them.
For example, in the box diagram Fig.~\ref{fig:BoxDiagram}(a)
the corresponding energy denominators are
\begin{eqnarray}
  &&\left(\frac{-i}{p_{1}^{-}+q_{1}^{-}-k_{1}^{-}-i\epsilon}\right),\ 
  \left(\frac{-i}{l_{1}^{-}+k_{2}^{-}-q_{1}^{-}-i\epsilon}\right),\ 
  \left(\frac{-i}{q_{2}^{-}-k_{2}^{-}-l_{2}^{-}-i\epsilon}\right),\nonumber\\
  &&\mbox{and}\;\;
  \left(\frac{-i}{k_{1}^{-}-q_{2}^{-}-p_{2}^{-}-i\epsilon}\right).
\end{eqnarray}
If we consider $q_{2}$ as the loop momentum,
by shifting momentum as
\begin{equation}
  k_{1}\rightarrow k_{1}+q_{2},\ 
  q_{1}\rightarrow q_{1}+q_{2}\;\;\mbox{and}\;\; 
  k_{2}\rightarrow k_{2}+q_{2},
\end{equation}
the denominators become
\begin{eqnarray}
  &&\left(\frac{-i}{p_{1}^{-}+q_{1}^{-}-k_{1}^{-}-i\epsilon}\right),\ 
  \left(\frac{-i}{l_{1}^{-}+k_{2}^{-}-q_{1}^{-}-i\epsilon}\right),\ 
  \left(\frac{-i}{-k_{2}^{-}-l_{2}^{-}-i\epsilon}\right),\nonumber\\
  &&\mbox{and}\;\;
  \left(\frac{-i}{k_{1}^{-}-p_{2}^{-}-i\epsilon}\right),
\end{eqnarray}
whose energies are assigned in Fig.~\ref{fig:BoxDiagram}(b).
Because the loop momenta are not restricted by the usual conservation
law of four-momenta in the covariant perturbation theory,
it is apparent that the energy denominators do not depend on them.

\figput{fig:BoxDiagram}

\subsection{Renormalization of $G$}\label{sec:RenormalizationG}
As mentioned in Sec.~\ref{sec:RuleOfG}, $\cA{n}$ corresponds to the
usual amplitude.
What is different from the amplitude is
that energies of the propagators in $\cA{n}$ are independent
each other.

Even if the energies are not conserved,
we can renormalize $\cA{n}$ by the usual prescription in the covariant
perturbation theory
because UV divergences from the integrations of the loop momenta are local.
They emerge as the coefficients of the polynomial of the other
momenta.
If the zeroth order term in those expansions
includes a divergence, it is renormalized by shifting masses
or a coupling constant.
The other divergences are logarithmic and depend on momenta,
then they must be renormalized by the field renormalization.

After renormalizing $\cA{n}$, we must consider the energy integrations.
First, 
we show that the new divergence arises
when we integrate the energies of the renormalized $\cA{n}$ 
with the energy denominators.
In Hamiltonian formalism,
it is not so trivial where the field renormalization constants come from
because the energies are not conserved.
We discuss the field renormalization 
and show that an ambiguity which may cancel the above divergence
comes from the energy integrations.

\subsubsection{Divergences in G}\label{sec:DivergenceG}
Even if $\cA{n}$ is renormalized, 
new divergences may arise from
the energy integrations in $\cV{n}$.

\figput{fig:OneParticleState}
This problem does not occur in the one-particle states
which correspond to the diagrams like in Fig.~\ref{fig:OneParticleState}.
From the on-shell condition
\begin{equation}
  p_{1}^{2}=p_{2}^{2}=m^{2},
\end{equation}
and the three-momentum conservation law
\begin{equation}
  p_{1}^{+}=p_{2}^{+},\quad p_{1\perp}=p_{2\perp},
\end{equation}
for the fermion line,
the one-particle states conserve the energies:
\begin{equation}
  p_{1}^{-}=\frac{p_{1\perp}^{2}+m^{2}}{2p_{1}^{+}}=
  \frac{p_{2\perp}^{2}+m^{2}}{2p_{2}^{+}}=p_{2}^{-}.
\end{equation}
The same conservation law is applied to the anti-fermion line.
It is obvious from Eq.~(\ref{eqn:S-G}) that
$G$ multiplied by the total energy conserving 
delta function is $S$.
Therefore the one-particle states of $G$ are equivalent to those of $S$.
If we renormalize the self-energy part under the physical renormalization
condition, $\cM{n}$ and $\cbM{n}$ in Eq.~(\ref{eqn:cGn}) vanish.
An example of the order $g^{2}$ will be demonstrated 
in Appendix~\ref{sec:NumericalResult}.

In multi-particle states, even though the external lines satisfy 
the on-shell condition and the three-momentum conservation law,
one can not say that the total energy is conserved.
We will show that such divergences disappear even in
the multi-particle states
except for the case that the outgoing external fermion line has
a self-energy part.

First, we consider the one-particle irreducible part
$\Gamma(k_{1}^{\prime},k_{2}^{\prime},\cdots)$ of the renormalized $\cA{n}$,
where $k_{1}^{\prime}, k_{2}^{\prime},\cdots$ are the four-momenta 
of the propagators
except for the loop momentum, and satisfy the three-momentum conservation law.
Since $\Gamma(k_{1}^{\prime},k_{2}^{\prime},\cdots)$ is covariant,
the analysis for large four-momentum $k_{i}^{\prime\mu}$'s,
which is used in the operator product expansion,
is valid when we investigate
the asymptotic behavior for large $k_{i}^{\prime -}$'s.\cite{SW} \ 
Therefore, after considering that
the loop momenta associated to $k_{i}^{\prime\mu}$'s
are as large as $k_{i}^{\prime\mu}$'s,
we regard $k_{i}^{\prime -}$'s as larger than 
$k_{i}^{\prime +}$'s and $k_{i\perp}^{\prime}$'s.
By the power counting,
the leading contribution comes from the self-energy part.
For the scalar self-energy part,
the leading behavior of $\Gamma(k_{1}^{\prime},k_{2}^{\prime},\cdots)$ goes
\begin{eqnarray}
  \Gamma(k_{1}^{\prime},k_{2}^{\prime},\cdots)_{\rm scalar\;self-energy}
  &\sim&
  \sum_{ij}c_{ij}\inp{k_{i}^{\prime}}{k_{j}^{\prime}}\nonumber\\
  &\sim&
  \sum_{ij}c_{ij}(k_{i}^{\prime +}k_{j}^{\prime -}
  +k_{i}^{\prime -}k_{j}^{\prime +}),
  \label{eqn:sse-asymptotic}
\end{eqnarray}
and for the fermion self-energy,
\begin{eqnarray}
  \Gamma(k_{1}^{\prime},k_{2}^{\prime},\cdots)_{\rm fermion\;self-energy}
  &\sim&
  \sum_{i}c_{i}\slk_{i}^{\prime},\nonumber\\
  &\sim&
  \sum_{i}c_{i}k_{i}^{\prime -}\gamma^{+},
  \label{eqn:fse-asymptotic}
\end{eqnarray}
where $c_{ij}$ and $c_{i}$ may depend on logarithmic factor of 
$k_{i}^{\prime}$'s.
It is important that both are proportional to
$k_{i}^{\prime -}$'s at most.
If we multiply $\Gamma(k_{1}^{\prime},k_{2}^{\prime},\cdots)$ by 
the energy denominators and integrate over the energies,
it seems that it behaves as logarithmically divergent.
But such term must not exist when the energies are conserved, that is,
such divergence is not possible in $S$, 
so it must be proportional to the total energy difference.
For the scalar self-energy in the internal line,
we regard the blob in Fig.~\ref{fig:SelfEnergy}(a) as 
$\Gamma(k_{1}^{\prime},k_{2}^{\prime},\cdots)_{\rm scalar\;self-energy}$.
The contribution from it is
\begin{eqnarray}
  &&\intenp{k_{1}}\intenp{k_{2}}\cdots
  \ed{q_{2}^{-}-k_{1}^{\prime -}-k_{2}^{\prime -}}\cdots
  \Gamma(k_{1}^{\prime},k_{2}^{\prime},\cdots)_{\rm scalar\;self-energy}
  \nonumber\\
  &\sim&\ed{q_{2}^{-}-q_{1}^{-}}
  \{i\Pi(q_{1},q_{2})+iq^{+}(q_{2}^{-}-q_{1}^{-})\log(q^{\prime +}\Lambda^{-})
  \nonumber\\
  &&\hspace{17em}+(q_{2}^{-}-q_{1}^{-})(\FTs)\},
  \label{eqn:sse-log}
\end{eqnarray}
where $\Pi(q_{1},q_{2})$ becomes the usual scalar self-energy 
if we replace the energy denominator
with the energy delta function, and $\Lambda^{-}$ is the UV cutoff of the
energy.
The second and third term are proportional to \mbox{$(q_{2}^{-}-q_{1}^{-})$}
by the above reason.
$q^{+}$ and $q^{\prime +}$ are proper longitudinal momenta.
The divergence only arises in the second term
because Eq.~(\ref{eqn:sse-log}) is regarded as Taylor expansion
in \mbox{$(q_{2}^{-}-q_{1}^{-})$}.
The first term is the ordinary renormalized finite term.
Although the second term is divergent,
it is found that it behaves as 
logarithmically divergent at most
by Lorentz covariance and the power counting .
The third term converges because the second term is logarithmically divergent
and a differentiation by the energy decreases the power of the divergence
by one.

Although Eq.~(\ref{eqn:sse-log}) includes the divergence,
it vanishes when it is inserted in the internal scalar line.
Let us consider scalar propagators and energy denominators in vertices
around the blob in Fig.~\ref{fig:SelfEnergy}(a).
It is crucial that 
the second term in Eq.~(\ref{eqn:sse-log}) does not depend 
$q_{1}^{-}$ and $q_{2}^{-}$.
The total contribution from the second term is
\begin{eqnarray}
  &&\inten{q_{1}}\inten{q_{2}}
  \ed{k_{4}^{-}-k_{3}^{-}-q_{2}^{-}}
  \DF{q_{2}}\nonumber\\
  &&\times
  \left\{\ed{q_{2}^{-}-q_{1}^{-}}iq^{+}(q_{2}^{-}-q_{1}^{-})
    \log(q^{\prime +}\Lambda^{-})\right\}
  \DF{q_{1}}
  \ed{q_{1}^{-}+k_{2}^{-}-k_{1}^{-}}\nonumber\\
  &=&\DF{k_{4}-k_{3}}\theta(-q_{2}^{+})
  q^{+}\log(q^{\prime +}\Lambda^{-})
  \DF{k_{1}-k_{2}}\theta(q_{1}^{+}).
  \label{eqn:s-div}
\end{eqnarray}
But the three-momenta are conserved 
($q_{1}^{+}=q_{2}^{+}, q_{1\perp}=q_{2\perp}$),
so it is obvious that Eq.~(\ref{eqn:s-div}) vanishes.

The fermion self-energy is logarithmically divergent 
like Eq.~(\ref{eqn:sse-log}).
For the internal fermion line in Fig.~\ref{fig:SelfEnergy}(b),
the contribution from the divergent part is
\begin{eqnarray}
  &&\inten{k_{2}}\inten{k_{3}}
  \ed{k_{4}^{-}+q_{2}^{-}-k_{3}^{-}}
  \SF{k_{3}}\nonumber\\
  &&\times\left\{\ed{k_{3}^{-}-k_{2}^{-}}i\gamma^{+}(k_{3}^{-}-k_{2}^{-})
    \log(k^{+}\Lambda^{-})\right\}
  \SF{k_{2}}
  \ed{k_{2}^{-}-k_{1}^{-}-q_{1}^{-}},
  \label{eqn:f-div}
\end{eqnarray}
where $k^{+}$ is a proper longitudinal momentum.
Unlike  $\DF{q}$, $\SF{k}$ has a term which does not depend on the energy
but is proportional to $\gamma^{+}$:
\begin{equation}
  \SF{k}=\frac{i\gamma^{+}}{2k^{+}}
  +i\frac{\frac{k_{\perp}^{2}+m^{2}}{2k^{+}}\gamma^{+}+k^{+}\gamma^{-}
    -k_{\perp}\cdot\gamma_{\perp}+m}{k^{2}-m^{2}+i\epsilon}.
\end{equation}
Fortunately, $\gamma^{+}$ has the property that $(\gamma^{+})^{2}=0$,
so the first term has no effect in Eq.~(\ref{eqn:f-div}).
The second term is discussed 
in the same way as in the scalar case, and then Eq.~(\ref{eqn:f-div}) vanishes.

Note that the above mechanism only works for the internal lines.
The diagrams which have the self-energy in the external fermion line
may have the same divergence.
We consider the incoming external fermion line like in 
Fig.~\ref{fig:SelfEnergy}(c).
The contribution from the divergent part becomes
\begin{eqnarray}
  &&\inten{k_{1}}\ed{q_{1}^{-}+k_{2}^{-}-k_{1}^{-}}\SF{k_{1}}
  \left\{\ed{k_{1}^{-}-p_{2}^{-}}i\gamma^{+}(k_{1}^{-}-p_{2}^{-})
    \log(k^{+}\Lambda^{-})\right\}u(p_{2},\lambda_{2})\nonumber\\
  &=&\SF{q_{1}+k_{2}}\theta(-p_{2}^{+})\log(k^{+}\Lambda^{-})
  u(p_{2},\lambda_{2}),
\end{eqnarray}
where all the three-momenta are conserved.
Since the incoming line has 
the positive longitudinal momentum $p_{2}^{+}>0$,
this vanishes.
In the case of the outgoing external fermion line like in 
Fig.~\ref{fig:SelfEnergy}(d),
the contribution from the divergent part becomes
\begin{eqnarray}
  &&\inten{k_{1}}\bar{u}(p_{1},\lambda_{1})
  \left\{\ed{p_{1}^{-}-k_{1}^{-}}i\gamma^{+}(p_{1}^{-}-k_{1}^{-})
    \log(k^{+}\Lambda^{-})\right\}
  \SF{k_{1}}\ed{k_{1}^{-}-k_{2}^{-}-q_{1}^{-}}\nonumber\\
  &=&\bar{u}(p_{1},\lambda_{1})\SF{k_{2}+q_{1}}\theta(p_{1}^{+})
  \log(k^{+}\Lambda^{-}).
\end{eqnarray}
Since the outgoing line satisfies $p_{1}^{+}>0$, this does not vanish
and is logarithmically divergent.

\figput{fig:SelfEnergy}

\subsubsection{Field renormalization}\label{sec:FieldRenormalization}
The counterterms for the fields have the same behaviors as those 
in Eqs.~(\ref{eqn:sse-asymptotic}) and (\ref{eqn:fse-asymptotic}).
The energy integrations of those yield not logarithmic divergences
but ambiguous constants.

First, let us consider the case of the scalar field.
We define the field renormalization constant as $\SZ$.
The corresponding counterterm in our rules is
\begin{equation}
  \mbox{Fig.~\ref{fig:ScalarFR}(a)}=\inten{k}
  \ed{q_{2}^{-}-k^{-}}
  (1-\SZ)\DFI{k}
  \ed{k^{-}-q_{1}^{-}},
\end{equation}
where the three-momenta are conserved (\mbox{$k_{1}^{+}=q_{1}^{+}=q_{2}^{+}$},
\mbox{$k_{1\perp}=q_{1\perp}=q_{2\perp}$}) and we omit
the scalar propagators attached on both sides of it.
This is the usual field renormalization
and cancels the logarithmic divergence which
comes from the integration of the loop momentum.
Carrying out the energy integration, we obtain
\begin{equation}
  (1-\SZ)\left[c q_{1}^{+}+\ed{q_{2}^{-}-q_{1}^{-}}\DFI{q_{1}}\right],
\end{equation}
where $c$ is an ambiguous constant which depends on the way of
taking the limit $k^{-}\rightarrow\pm\infty$.
More generally, the contribution from $n$ counterterms connected 
by the scalar propagators is
\begin{eqnarray}
  \mbox{Fig.~\ref{fig:ScalarFR}(b)}
  &=&\inten{k_{1}}\inten{k_{2}}\cdots\inten{k_{n}}
  \inten{q_{2}}\inten{q_{3}}\cdots\inten{q_{n}}\nonumber\\
  &&\ed{q_{n+1}^{-}-k_{n}^{-}}
  (1-\SZ)\DFI{k_{n}}
  \ed{k_{n}^{-}-q_{n}^{-}}
  \cdots\nonumber\\
  &&\hspace{-3em}\cdots
  (1-\SZ)\DFI{k_{2}}
  \ed{k_{2}^{-}-q_{2}^{-}}
  \DF{q_{2}}
  \ed{q_{2}^{-}-k_{1}^{-}}
  (1-\SZ)\DFI{k_{1}}
  \ed{k_{1}^{-}-q_{1}^{-}}\nonumber\\
  &=&(1-\SZ)^{n}\left[s_{n}(q_{1}^{+})q_{1}^{+}+\ed{q_{n+1}^{-}-q_{1}^{-}}
    \DFI{q_{1}}\right],
  \label{eqn:IncAmbiguity}
\end{eqnarray}
where all the three-momenta are conserved.
$s_{n}(q^{+})$ is defined as
\begin{equation}
  s_{n}(q^{+})=a_{n}\theta(q^{+})+b_{n}\theta(-q^{+}),
  \label{eqn:sn}
\end{equation}
where  $a_{n}$ and $b_{n}$ are ambiguous constants
because there is an ambiguity
of the order of the energy 
integrations.\footnote{See Appendix~\ref{sec:Ambiguity}.} \ 
If we insert it in an internal scalar line,
the contribution becomes
\begin{eqnarray}
  \mbox{Fig.~\ref{fig:ScalarFR}(c)}  
  &=&\inten{q_{1}}\inten{q_{n+1}}
  \ed{k_{4}^{-}-k_{3}^{-}-q_{n+1}^{-}}\DF{q_{n+1}}
  \nonumber\\
  &&\times
  (1-\SZ)^{n}\left[s_{n}(q_{1}^{+})q_{1}^{+}+\ed{q_{n+1}^{-}-q_{1}^{-}}
    \DFI{q_{1}}\right]\nonumber\\
  &&\times\DF{q_{1}}
  \ed{q_{1}^{-}+k_{2}^{-}-k_{1}^{-}}\nonumber\\
  &=&\DF{k_{4}-k_{3}}\theta(-q_{n+1}^{+})
  (1-\SZ)^{n}s_{n}(q_{1}^{+})q_{1}^{+}
  \DF{k_{1}-k_{2}}\theta(q_{1}^{+})\nonumber\\
  &&+\inten{q}
  \ed{k_{4}^{-}-k_{3}^{-}-q^{-}}(1-\SZ)^{n}\DF{q}
  \ed{q^{-}+k_{2}^{-}-k_{1}^{-}}\nonumber\\
  &=&\inten{q}
  \ed{k_{4}^{-}-k_{3}^{-}-q^{-}}(1-\SZ)^{n}\DF{q}
  \ed{q^{-}+k_{2}^{-}-k_{1}^{-}}.
\end{eqnarray}
This does not depend on $s_{n}(q^{+})$ and
is equivalent to a free propagator multiplied 
by the constant $(1-\SZ)^{n}$.
The mechanism that $s_{n}(q^{+})$ vanishes
is the same as one that the logarithmic divergence in the self-energy does.
We take the sum of the diagrams like in Fig.~\ref{fig:ScalarFR}(d)
and obtain the result
\begin{eqnarray}
  &&\left(1+\sum_{n=1}^{\infty}(1-\SZ)^{n}\right)
  \inten{q}\ed{k_{4}^{-}-k_{3}^{-}-q^{-}}\DF{q}
  \ed{q^{-}+k_{2}^{-}-k_{1}^{-}}\nonumber\\
  &=&\frac{1}{\SZ}
  \inten{q}\ed{k_{4}^{-}-k_{3}^{-}-q^{-}}\DF{q}
  \ed{q^{-}+k_{2}^{-}-k_{1}^{-}},
  \label{eqn:s-internal}
\end{eqnarray}
where the three-momenta are conserved.
This property is the same as that of the usual field renormalization.

\figput{fig:ScalarFR}

Similarly, we can discuss the fermion field.
If we connect $n$ counterterms for the fermion field with fermion propagators,
the contribution becomes
\begin{eqnarray}
  \mbox{Fig.~\ref{fig:FermionFR}(a)}
  &=&\inten{q_{1}}\inten{q_{2}}\cdots\inten{q_{n}}
  \inten{k_{2}}\inten{k_{3}}\cdots\inten{k_{n}}\nonumber\\
  &&\ed{k_{n+1}^{-}-q_{n}^{-}}
  (1-\FZ)\SFI{q_{n}}
  \ed{q_{n}^{-}-k_{n}^{-}}
  \cdots\nonumber\\
  &&\hspace{-3em}\cdots
  (1-\FZ)\SFI{q_{2}}
  \ed{q_{2}^{-}-k_{2}^{-}}
  \SF{k_{2}}
  \ed{k_{2}^{-}-q_{1}^{-}}
  (1-\FZ)\SFI{q_{1}}
  \ed{q_{1}^{-}-k_{1}^{-}}\nonumber\\
  &=&(1-\FZ)^{n}\left[f_{n}(k_{1}^{+})\gamma^{+}+\ed{k_{n+1}^{-}-k_{1}^{-}}
    \SFI{k_{1}}\right],
\end{eqnarray}
where all the three-momenta are conserved and $\FZ$ is 
the field renormalization constant for the fermion field.
The function $f_{n}(k^{+})$ is defined as
\begin{equation}
  f_{n}(k^{+})=c_{n}\theta(k^{+})+d_{n}\theta(-k^{+}),
\end{equation}
where $c_{n}$ and $d_{n}$ are ambiguous constants
due to a similar ambiguity.

When we insert Fig.~\ref{fig:FermionFR}(a) into the internal fermion line,
the contribution from it is given by
\begin{eqnarray}
  \mbox{Fig.~\ref{fig:FermionFR}(b)}
  &=&\inten{k_{1}}\inten{k_{n+1}}\ed{q_{2}^{-}+p_{2}^{-}-k_{n+1}^{-}}
  \SF{k_{n+1}}\nonumber\\
  &&\times (1-\FZ)^{n}
  \left[f_{n}(k_{1}^{+})\gamma^{+}+\ed{k_{n+1}^{-}-k_{1}^{-}}\SFI{k_{1}}
  \right]\nonumber\\
  &&\times\SF{k_{1}}\ed{k_{1}^{-}-q_{1}^{-}-p_{1}^{-}}\nonumber\\
  &=&\SF{q_{2}+p_{2}}\theta(-k_{n+1}^{+})(1-\FZ)^{n}f_{n}(k_{1}^{+})\gamma^{+}
  \SF{q_{1}+p_{1}}\theta(k_{1}^{+})\nonumber\\
  &&+\inten{k}\ed{q_{2}^{-}+p_{2}^{-}-k^{-}}(1-\FZ)^{n}\SF{k}
  \ed{k^{-}-q_{1}^{-}-p_{1}^{-}}\nonumber\\
  &=&\inten{k}\ed{q_{2}^{-}+p_{2}^{-}-k^{-}}(1-\FZ)^{n}\SF{k}
  \ed{k^{-}-q_{1}^{-}-p_{1}^{-}},
\end{eqnarray}
which is a free propagator multiplied by the constant $(1-\FZ)^{n}$
and the energy denominators.
An infinite sum of the diagrams becomes
\begin{equation}
  \frac{1}{\FZ}\inten{k}\ed{q_{2}^{-}+p_{2}^{-}-k^{-}}\SF{k}
  \ed{k^{-}-q_{1}^{-}-p_{1}^{-}}\label{eqn:f-internal}.
\end{equation}
As a result, $f_{n}(k^{+})$ does not affect the fermion propagator.

For the incoming external fermion line, the contribution from
the $n$ connected counterterms is
\begin{eqnarray}
  \mbox{Fig.~\ref{fig:FermionFR}(c)}
  &=&(1-\FZ)^{n}[f_{n}(k_{1}^{+})\theta(-k_{1}^{+})+1]
  u(k_{1},\lambda)\ed{q_{1}^{-}+p_{1}^{-}-k_{1}^{-}}\nonumber\\
  &=&(1-\FZ)^{n}u(k_{1},\lambda)\ed{q_{1}^{-}+p_{1}^{-}-k_{1}^{-}},
\end{eqnarray}
where $k_{1}^{2}=m^{2}$ and $k_{1}^{+}>0$.
Of course, all the three-momenta are conserved.
In this case, $f_{n}(k^{+})$ does not affect it.
The infinite sum becomes
\begin{equation}
  \frac{1}{\FZ}u(k,\lambda)\ed{q_{1}^{-}+p_{1}^{-}-k^{-}},
  \label{eqn:f-in}
\end{equation}
which is $1/\FZ$ times the tree external line.
$k^{\mu}$ is the momentum of the external fermion.

For the outgoing external fermion line, the contribution is
\begin{eqnarray}
  \mbox{Fig.~\ref{fig:FermionFR}(d)}
  &=&(1-\FZ)^{n}[f_{n}(k_{n+1}^{+})\theta(k_{n+1}^{+})+1]
  \bar{u}(k_{n+1},\lambda)
  \ed{k_{n+1}^{-}-q_{1}^{-}-p_{1}^{-}}\nonumber\\
  &=&(1-\FZ)^{n}(c_{n}+1)\bar{u}(k_{n+1},\lambda)
  \ed{k_{n+1}^{-}-q_{1}^{-}-p_{1}^{-}},
  \label{eqn:f-exto}
\end{eqnarray}
where 
$k_{n+1}^{2}=m^{2}$ and $k_{n+1}^{+}>0$.
The infinite sum becomes
\begin{equation}
  \left\{1+\sum_{n=1}^{\infty}(1-\FZ)^{n}(c_{n}+1)\right\}\bar{u}(k,\lambda)
  \ed{k^{-}-q_{1}^{-}-p_{1}^{-}},
  \label{eqn:f-out}
\end{equation}
which is the tree external line multiplied by the constant.
Note that $c_{n}$ may not be equal to zero.

\figput{fig:FermionFR}

\subsubsection{Summary of the renormalization of $G$}\label{sec:SubSummary}
We have shown that 
UV divergences from the loop integrations are renormalized by the usual
procedure in the covariant perturbation theory.
After renormalizing $\cA{n}$, we carry out the energy integrations.
As a result, the new UV divergence and ambiguous constant $c_{n}$
arise only from the diagrams that the fermion self-energy and its counterterm
are inserted into the outgoing external line.

We fix $c_{n}$ so that it can remove the above divergence.
This is always possible because both live in the same place and
are closely related.
Although it is not the only way to fix $c_{n}$,
we consider that it is natural for both to cancel each other.

\subsection{Divergences in $\Heff$}\label{sec:DivergenceHeff}
In Sec.~\ref{sec:RenormalizationG}, we have shown that $G$ is not divergent
after renormalizing it by the usual prescription in the usual
perturbation theory.
In this subsection, we show that $\Heff$ is divergent even if $G$ is finite.

From Eqs.~(\ref{eqn:FinalEffHamiltonian}), (\ref{eqn:cF2}),
and (\ref{eqn:cF4}),
the effective Hamiltonian is written as
\begin{eqnarray}
  \Heff&=&\frac{1}{2}\eta\left\{H_{0}+\cG{4}
    +i\cG{2}\int_{-\infty}^{0}dx^{+}\cG{2}(x^{+})\right.\nonumber\\
  &&\hspace{2em}\left.+\frac{i}{4}\int_{-\infty}^{0}dx^{+}
    [\cG{2}(x^{+})-\cG{2}^{\dag}(x^{+}),\cG{2}]\right\}\eta+\HC+\cO(g^{6}),
  \label{eqn:EffHamiltonian-G}
\end{eqnarray}
to the order of $g^{4}$.
We can renormalize $\cG{2}$ and $\cG{4}$, but new divergences arise
from the product of $\cG{2}$.
One of the examples is a product of two one-scalar-exchange diagrams 
in Fig.~\ref{fig:Exchange}.
The corresponding term in $\cG{2}$ is
\begin{eqnarray}
  \cG{2}^{\rm ex}&=&
  -i\sum_{\lambda_{1}}\sum_{\lambda_{2}}\sum_{\sigma_{1}}\sum_{\sigma_{2}}
  \int_{p_{1}}\int_{p_{2}}\int_{l_{1}}\int_{l_{2}}
  \bar{u}(p_{1},\lambda_{1})(-g\gamma_{5})u(p_{2},\lambda_{2})
  \bar{v}(l_{2},\sigma_{2})(-g\gamma_{5})v(l_{1},\sigma_{1})
  \nonumber\\
  &&\times
  [\DF{l_{2}-l_{1}}\theta(l_{2}^{+}-l_{1}^{+})+\DF{p_{1}-p_{2}}\theta(p_{2}^{+}-p_{1}^{+})]\nonumber\\
  &&\times \eta b^{\dag}_{\lambda_{1}}(p_{1})d^{\dag}_{\sigma_{1}}(l_{1})
  d_{\sigma_{2}}(l_{2})b_{\lambda_{2}}(p_{2})\eta
  (2\pi)^{3}\delta^{3}(p_{1}+l_{1}-p_{2}-l_{2}),
\end{eqnarray}
where we have integrated the energy of $\DF{q}$.
The product of two $\cG{2}^{\rm ex}$'s 
included in the third term of Eq.~(\ref{eqn:EffHamiltonian-G}) is
\begin{eqnarray}
  &&i\cG{2}^{\rm ex}\int_{-\infty}^{0}dx^{+}\cG{2}^{\rm ex}(x^{+})
  \nonumber\\
  &=&
  \frac{g^{4}}{16\pi^{2}}
  \log\Lambda_{\perp}^{2}
  \sum_{\lambda}\sum_{\sigma}
  \int_{p_{1}}\int_{p_{2}}\int_{l_{1}}\int_{l_{2}}
  \sqrt{x(1-x)x^{\prime}(1-x^{\prime})}
  \nonumber\\
  &&\times
  \left\{\frac{\theta(x-x^{\prime})}{1-x^{\prime}}
    \left(\frac{\log\frac{x}{x^{\prime}}}{1-x}+\frac{1}{x}\right)
    +
    \frac{\theta(x^{\prime}-x)}{x^{\prime}}
    \left(\frac{\log\frac{1-x}{1-x^{\prime}}}{x}+\frac{1}{1-x}\right)
  \right\}\nonumber\\
  &&\times \eta b^{\dag}_{\lambda}(p_{1})d^{\dag}_{\sigma}(l_{1})
  d_{\sigma}(l_{2})b_{\lambda}(p_{2})\eta
  (2\pi)^{3}\delta^{3}(p_{1}+l_{1}-p_{2}-l_{2})+(\FTs),
  \label{eqn:cF4-div}
\end{eqnarray}
\figput{fig:Exchange}
where $\Lambda_{\perp}$ is the cutoff of the transverse momentum of the
external fermion.
$x$ and $x^{\prime}$ are defined as
\begin{equation}
  x=\frac{p_{1}^{+}}{p_{1}^{+}+l_{1}^{+}},
  x^{\prime}=\frac{p_{2}^{+}}{p_{2}^{+}+l_{2}^{+}},
\end{equation}
respectively.
Eq.~(\ref{eqn:cF4-div}) is logarithmically divergent as 
$\Lambda_{\perp}\rightarrow\infty$.
A Feynman box diagram, which is finite by the power counting,
consists of the sum of the various
time-ordered diagrams which may be divergent individually.
The above product is one of such and is logarithmically divergent.

It is important to recognize that $\Heff$ should be divergent.
The reason is understood by the perturbative diagonalization of it:
\begin{eqnarray}
  E_{\rm pert}
  &=&\frac{1}{2}\langle\alpha|
  \left\{H_{0}+\cG{4}\frac{\mathstrut}{\mathstrut}
  \right.\nonumber\\
  &&\left.+\frac{i}{2}\cG{2}\int_{-\infty}^{0}dx^{+}\cG{2}(x^{+})
    +\frac{i}{2}\int_{-\infty}^{0}dx^{+}\cG{2}(x^{+})\cG{2}+\HC\right\}
  |\alpha\rangle
  +\cO(g^{6}),
\end{eqnarray}
where $|\alpha\rangle$ is an eigenstate of $H_{0}$.
Because the logarithmic divergence in Eq.~(\ref{eqn:cF4-div}) comes
from the integral of the three-momentum of the intermediate state,
the divergences of $\frac{i}{2}\int_{-\infty}^{0}dx^{+}\cG{2}(x^{+})\cG{2}$
and of $\frac{i}{2}\cG{2}\int_{-\infty}^{0}dx^{+}\cG{2}(x^{+})$
have the opposite sign,
and they cancel each other.
In general, using the eigenstates of $H_{0}$ and its complete set,
we obtain the matrix element
\begin{eqnarray}
  &&\langle\alpha|
  \left\{ \frac{i}{2}\cG{2}\int_{-\infty}^{0}dx^{+}\cG{2}(x^{+})
    +\frac{i}{2}\int_{-\infty}^{0}dx^{+}\cG{2}(x^{+})\cG{2}\right\}
  |\beta\rangle
  \nonumber\\
  &=&\sum_{\gamma}\langle\alpha|
  \cG{2}|\gamma\rangle\langle\gamma|\cG{2}|\beta\rangle
  \frac{1}{2}\left(\frac{1}{p_{\gamma}^{-}-p_{\beta}^{-}-i\epsilon}
    +\frac{1}{p_{\alpha}^{-}-p_{\gamma}^{-}-i\epsilon}\right)\nonumber\\
  &=&\sum_{\gamma}\langle\alpha|
  \cG{2}|\gamma\rangle\langle\gamma|\cG{2}|\beta\rangle
  \frac{1}{2}\frac{(p_{\alpha}^{-}-p_{\beta}^{-})}
  {(p_{\gamma}^{-}-p_{\beta}^{-}-i\epsilon)
    (p_{\alpha}^{-}-p_{\gamma}^{-}-i\epsilon)},
  \label{eqn:Cancellation}
\end{eqnarray}
in which the power of the intermediate energy $p_{\gamma}^{-}$ is decreased
by one, and thus
the power of the transverse momentum of the intermediate states by two.

It is important that although these divergences 
must be canceled by adding the artificial counterterms
in the other similarity methods,
they will {\em automatically} arise in the higher order terms in our method.
As shown in Appendix~\ref{sec:NumericalResult},
if we diagonalize the effective Hamiltonian to the order of $g^{2}$,
the eigenvalue is divergent.
In general, such divergences are
mainly related to the box diagrams 
and arise
not only in this case but also in the case of TD approximation in
$(3+1)$ dimension.\cite{G-H-P-S-W} \ 
Eq.~(\ref{eqn:Cancellation}) says that 
such divergences are canceled if we include $g^{4}$ order terms,
then the divergences in the eigenvalue are weakened.
However, new divergences arise in diagonalizing the effective Hamiltonian
due to new interactions.
We expect that they are canceled by the higher order interactions
because the exact eigenvalue should not depend on the cutoff
and the similarity transformation does not change the eigenvalue.

\section{Summary and discussions}\label{sec:Summary}
In this paper, we have shown that the effective Hamiltonian, which is obtained 
by the FSTO's similarity transformation in the particle number space, can be
written in terms only of $G$ (or $F$)
to the fourth order in $H^{\prime}$.

To introduce $G$ is crucial for constructing 
the effective Hamiltonian
more easily than in the traditional way especially in the higher orders.
$G$ has a favorable property that it is diagonal in the particle number space.
Since it also
has the same form of the formula of the S-matrix operator,
we can use the Feynman diagrams and the rules mentioned
in Sec.~\ref{sec:RuleOfG}
in the LF Yukawa model.
By using the knowledge in the covariant perturbation theory,
we can avoid the complexity in calculating it
and make the renormalization procedure transparent.
Note that our construction rules are a little bit different from the usual
Feynman rules
because energies are not conserved in each vertices.

The divergences due to the integrations of the loop momenta are renormalized
by the familiar prescription in the covariant perturbation theory.
The divergences due to the energy integrations can be canceled by terms which
come from the ambiguity of the counterterms for the fields.

The mechanism of the cancellation of the extra divergences
from the energy integrations
is valid only in the LF field theory.
It will be applied not only to the LF Yukawa model
but also to the other LF models.
Although we can not use it in the ET field theory,
we expect that
more precise integrations and
treatment of the limit
are necessary for the cancellation of the divergences.

Although $G$ is constructed by Feynman diagrams and renormalized,
the effective Hamiltonian has divergent terms
which are written in terms of products of the renormalized those.
It is very important that such terms do not need the counterterms
but work as those which cancel the divergences in diagonalizing 
the effective Hamiltonian.
Although the exact eigenvalue should not depend on the cutoff,
our method is perturbative, 
so it is not possible to treat
nonperturbative divergences in diagonalization and to get the exact eigenvalue.
It is highly desired to find the nonperturbative counterterms for it.
When we find the low energy states,
our method is useful in small coupling region enough to
allow one to ignore the cutoff dependence
if the cutoff is much larger than lower eigenvalues.

There is also the problem of vanishing energy denominator.
Here, we only consider the LF Yukawa model without a massless particle,
so that we may avoid this problem by $-i\epsilon$ in denominators.
If there is a massless particle like in QCD, we must be attentive 
to it.

We proved here that the effective Hamiltonian can be
written in terms of $G$ to the fourth order by
the explicit calculation.
But the general proof for higher orders is lacking.
Although we do not know how the general proof goes, we think that it is likely
that this feature persists to all orders.
Recently Hansper proposed a nonperturbative approach for the FSTO's method
in LF field theory.\cite{H} \ 
Although he applied it only to the parton distributions,
it is very useful to solve the mesonic
bound states in QCD
if it is applicable to our method.

In Appendix~\ref{sec:NumericalResult},
when the coupling constant is large enough,
the lowest-energy state
has the positive binding energy 
in the numerical calculation,
but we do not regard it as a bound state 
because it depends on
the transverse cutoff $\Lambda_{\perp}$.
As mentioned above, our method is valid in small coupling
and we expect that it is improved in higher order calculations.
We are now extending the present work to the next order
in order to confirm the cancellation of the cutoff dependence and
to get bound states.\cite{Y} \ 

It is interesting to note that this method might also be applied 
to the similarity transformation in momentum space because
Eq.~(\ref{eqn:FinalEffHamiltonian}) does not depend on the choice of $\eta$.
If so, to get the effective Hamiltonian would be much easier than
the traditional way.

\section*{Acknowledgement}
The author is grateful to K. Harada for helpful discussions and encouragement,
and also thanks A. Okazaki for his useful work.

\appendix
\section{Light-Front Yukawa model}\label{sec:LF_Yukawa}
In the LF framework, we use $x^{+}=(x^{0}+x^{3})/\sqrt{2}$ as a time
variable instead of $t$.
The Lagrangian of the this model is
\begin{equation}
  {\cal L}=\frac{1}{2}\partial_{\mu}\phi\,\partial^{\mu}\phi
  -\frac{1}{2}\mu^{2}\phi^{2}
  +\bar{\psi}
  (i\gamma^{\mu}\partial_{\mu}-m+ig\gamma_{5}\phi)
  \psi.
  \label{eqn:Lagrangian}
\end{equation}
By LF quantization with Dirac's quantization method,
dynamical fields must satisfy the following commutation relations
\begin{equation}
  [\,\phi(x),\partial_{-}\phi(y)\,]_{x^{+}=y^{+}}
  =\frac{i}{2}\delta(x^{-}-y^{-})\delta^{2}(x_{\perp}-y_{\perp}),
\end{equation}
\begin{equation}
  \{\xi(x),\xi^{\dag}(y)\}_{x^{+}=y^{+}}
  =\frac{1}{\sqrt{2}}\delta(x^{-}-y^{-})\delta^{2}(x_{\perp}-y_{\perp})
  \Lambda_{+},
\end{equation}
where $\xi(x)$ is the dynamical part of the fermion field $\psi(x)$:
\begin{equation}
  \xi(x)=\Lambda_{+}\psi(x).
\end{equation}
$\Lambda_{\pm}$ are the projection operators of the fermion field and defined
as
\begin{equation}
  \Lambda_{\pm}=\frac{1}{\sqrt{2}}\gamma^{0}\gamma^{\pm}.
\end{equation}
From Eq.~(\ref{eqn:Lagrangian}), 
the familiar Legendre transformation gives the LF Hamiltonian
\begin{equation}
  H=P^{-}=H_{0}+\Hp,
\end{equation}
where
\begin{eqnarray}
  H_{0}&=&\int dx^{-}d^{2}x_{\perp}\left\{
    \frac{1}{2}(\partial_{\perp}\phi)^{2}+\frac{1}{2}\mu^{2}\phi^{2}
    +\xi^{\dag}\frac{-\partial_{\perp}^{2}+m^{2}}{\sqrt{2}i\partial_{-}}\xi
  \right\},\\
\Hp&=&\int dx^{-}d^{2}x_{\perp}\left\{
  -ig\left[\xi^{\dag}\gamma^{0}\gamma_{5}\phi\frac{1}
    {\sqrt{2}i\partial_{-}}
    (i\inp{\alpha_{\perp}}{\partial_{\perp}}+m\beta)\xi
    -\HC\right]\right.\nonumber\\
  &&\hspace{5em}\left.+\frac{g^{2}}{2}\left[(\xi^{\dag}\phi)
      \frac{1}{\sqrt{2}i\partial_{-}}(\phi\xi)
      -\frac{1}{\sqrt{2}i\partial_{-}}(\phi\xi^{T})
      (\xi^{*}\phi)\right]\right\},
\end{eqnarray}
and
\begin{equation}
  \alpha_{\perp}=\gamma^{0}\gamma_{\perp},\;\;
  \beta=\gamma^{0}.
\end{equation}

In the interaction picture, they become
\begin{eqnarray}
  H_{0}(x^{+}) & = & \int dx^{-}d^{2}x_{\perp}
  \left\{
    \frac{1}{2}(\partial_{\perp}\phi)^{2}
    +\frac{1}{2}\mu^{2}\phi^{2}
    +\bar{\psi}\gamma^{+}
    \frac{-\partial_{\perp}^{2}+m^{2}}
    {2i\partial_{-}}\psi
  \right\},\\
  \HIp(x^{+})&=&\int dx^{-}d^{2}x_{\perp}
  \left\{-ig\bar{\psi}\gamma_{5}\phi\psi
  \frac{\mathstrut}{\mathstrut}
  +\frac{g^{2}}{2}\left[\bar{\psi}\phi
      \frac{\gamma^{+}}{2i\partial_{-}}(\phi\psi)
      -(\frac{\gamma^{+}}{2i\partial_{-}}\phi\psi)^{\rm T}
      (\bar{\psi}\phi)^{\rm T}\right]
  \right\}\nonumber\\
  &&\hspace{5em}+H_{\rm CT}(x^{+}),\frac{\mathstrut}{\mathstrut}
\end{eqnarray}
where all the field operators are defined in this picture.
Note that $\psi(x)$ is not the one in Eq.~(\ref{eqn:Lagrangian})
but a new field made from $\xi(x)$ in this picture,
and the solution of the free Dirac equation:\cite{B-P-P} \ 
\begin{equation}
  \psi(x)=\left[1+\frac{1}{\sqrt{2}i\partial_{-}}
    \left(\inp{i\alpha_{\perp}}{\partial_{\perp}}+m\beta\right)
  \right]\xi(x).
\end{equation}
$H_{\rm CT}(x^{+})$ is the part which corresponds to the usual counterterms
obtained by shifting the parameters.

We assume that the LF diagrams are equivalent to the covariant ones.
More concretely, in the LF framework the fermion propagator 
\begin{equation}
  \SF{k}=\int d^{4}x \langle 0|
  {\rm T^{+}}(\psi(x)\bar{\psi}(0)) |0\rangle e^{ik\cdot x}
  =\frac{i}{\slk-m+i\epsilon}
  -\frac{i}{2k^{+}}\gamma^{+},
  \label{eqn:FermionPropagator}
\end{equation}
includes the noncovariant term.
But in the first two diagrams in Fig.~\ref{fig:Instantaneous},
the contributions from the second term
are naively canceled by the vertex
\begin{equation}
  -ig^{2}\gamma^{+}\left\{\frac{1}{2(k_{1}^{+}+q_{1}^{+})}
    +\frac{1}{2(k_{1}^{+}-q_{2}^{+})}\right\}
  \label{eqn:Instantaneous}
\end{equation}
in the last diagram in Fig.~\ref{fig:Instantaneous},
which comes from the instantaneous interaction in the Hamiltonian.
Therefore, we regard the fermion propagator as the first term in 
Eq.~(\ref{eqn:FermionPropagator}) and covariant effectively,
and omit the second term in Eq.~(\ref{eqn:FermionPropagator})
and the vertex Eq.~(\ref{eqn:Instantaneous}) together.
As a result, the scalar and fermion propagator are given by
\begin{equation}
  \DF{q}=\frac{i}{q^{2}-\mu^{2}+i\epsilon},
\end{equation}
\begin{equation}
  \SF{k}=\frac{i}{\slk-m+i\epsilon}.
\end{equation}

\figput{fig:Instantaneous}

\section{Explicit calculations and numerical result}\label{sec:NumericalResult}
We calculate the ground state energy numerically to the second order in $g$.
Of course, the calculations to this order
are not so different from the other methods;\cite{PhD-O} \ 
the advantage of the present formulation becomes apparent
in the higher orders.\cite{Y}. \ 
The reason why we present the second order calculation here is to demonstrate
some features of our method and to clarify what would be expected in the
next order.

From Eqs.~(\ref{eqn:FinalEffHamiltonian}) and (\ref{eqn:cF2}),
the LF effective Hamiltonian is
\begin{equation}
  \Heff=
  \frac{1}{2}\eta\left(H_{0}+\cG{2}\right)\eta+\HC+\cO(g^{4})
  \label{eqn:LF_EffHamiltonian}
\end{equation}
to $g^{2}$ order.
Also, we add the flavor of fermions.
It is easy to estimate Eq.~(\ref{eqn:LF_EffHamiltonian}) from
Feynman diagrams by using our rules.

Since possible graphs are the same as those needed in constructing 
the S-matrix,
we can immediately imagine those which contribute to 
Eq.~(\ref{eqn:LF_EffHamiltonian}).
The Feynman diagrams associated with the effective Hamiltonian to the
second order are shown in
Fig.~\ref{fig:SecondOrder}.

\figput{fig:SecondOrder}

Terms which should be renormalized are only fermion self-energy terms 
$\cM{2}$ and $\cbM{2}$ in this order.
Although we do not give the rules for those,
they are much the same as those for $\cV{n}$.
The self-energy terms correspond to the first and the third graphs
in the second line of Fig.~\ref{fig:SecondOrder} and are written as
\begin{eqnarray}
  \cG{2}^{\rm self}&=-&\sum_{\lambda}\int_{p_{1}}\int_{p_{2}}
  (p_{1}^{-}-p_{2}^{-})
  \inten{k}
  \left(\ed{p_{1}^{-}-k^{-}}\right)
  \left(\ed{k^{-}-p_{2}^{-}}\right)\nonumber\\
  &&\times\bar{u_{i}}(p_{1},\lambda)
  \left\{
    \int\frac{d^{4}q}{(2\pi)^{4}}
    \DF{q}(-g\gamma_{5})\SFF{i}{k-q}(-g\gamma_{5})\right.\nonumber\\
  &&\hspace{5em}\left.\frac{\mathstrut}{\mathstrut}+
    \left[(1-{\FZ}_{i})^{(2)}\SFFI{i}{k}-i\delta m_{i}^{(2)}\right]
  \right\}u_{i}(p_{2},\lambda)\nonumber\\
  &&\times \eta b_{i\lambda}^{\dag}(p_{1})b_{i\lambda}(p_{2})\eta
  (2\pi)^{3}\delta(p_{1}-p_{2}),
\end{eqnarray}
where
$\delta m_{i}^{(2)}$'s and $(1-{\FZ}_{i})^{(2)}$'s 
are the masses and field renormalization constants of
order $g^{2}$ respectively,
and the on-shell condition $(p_{1}^{2}=p_{2}^{2}=m_{i}^{2})$
for the external fermion lines is satisfied.
Although there is no energy conserving delta function,
$p_{1}^{-}=p_{2}^{-}$ is satisfied
due to $p_{1}^{+}=p_{2}^{+}$, $p_{1\perp}=p_{2\perp}$ and 
the on-shell condition.
Therefore
\begin{eqnarray}
  &&-(p_{1}^{-}-p_{2}^{-})
  \left(\ed{p_{1}^{-}-k^{-}}\right)\left(\ed{k^{-}-p_{2}^{-}}\right)\nonumber\\
  &=&\frac{1}{p_{1}^{-}-k^{-}-i\epsilon}
  +\frac{1}{k^{-}-p_{1}^{-}-i\epsilon}\nonumber\\
  &=&2{\pi}i\delta(p_{1}^{-}-k^{-}).
\end{eqnarray}
This is what is said in the beginning of Sec.~\ref{sec:DivergenceG}.
Integrating $k^{-}$, we obtain
\begin{equation}
  \cG{2}^{\rm self}
  =\sum_{i}\sum_{\lambda}\int_{p}
  \frac{1}{2p^{+}}\bar{u_{i}}(p,\lambda)\Sigma_{i}^{(2)}(p)u_{i}(p,\lambda)
  \;\eta b_{i\lambda}^{\dag}(p)b_{i\lambda}(p)\eta,
\end{equation}
where
\begin{equation}
  \Sigma_{i}^{(2)}(p)=
  i\int{\frac{d^{4}q}{(2\pi)^{4}}
    \DF{q}
    (-g\gamma_{5})
    \SFF{i}{p-q}
    (-g\gamma_{5})}
  +\delta m_{i}^{(2)},
  \label{eqn:Sigma}
\end{equation}
and $p^{2}=m_{i}^{2}$.
The counterterm for the fermion field in Eq.~(\ref{eqn:Sigma})
vanishes by using the Dirac equation.
Obviously, $\Sigma_{i}^{(2)}(p)$ is the same as fermion self-energy terms
in the covariant perturbation theory.
It is convenient to impose the physical renormalization condition
\begin{equation}
  \bar{u}_{i}(p,\lambda)\Sigma_{i}^{(2)}(p)u_{i}(p,\lambda)
=0 \;\;\;\;(p^{2}=m_{i}^{2}),
\end{equation}
in order for the fermion masses to be the physical one.
For anti-fermion, the same argument is applied.

The rest are the scalar exchange part and the fermion
annihilation part which are represented by the last two graphs
in Fig.~\ref{fig:SecondOrder}.
After integrating the energies, we obtain
\begin{eqnarray}
  \cG{2}^{\rm ex}+\cG{2}^{{\rm ex}\dag}
  &=&g^{2}
  \sum_{\lambda_{1}}\sum_{\lambda_{2}}\sum_{\sigma_{1}}\sum_{\sigma_{2}}
  \int_{p_{1}}\int_{p_{2}}\int_{l_{1}}\int_{l_{2}}
  (2\pi)^{3}\delta^{3}(p_{1}+l_{1}-p_{2}-l_{2})\nonumber\\
  &&\hspace{1em}
  \times\bar{u}_{i}(p_{1},\lambda_{1})\gamma_{5}u_{i}(p_{2},\lambda_{2})
  \bar{v}_{j}(l_{2},\sigma_{2})\gamma_{5}v_{j}(l_{1},\sigma_{1})\nonumber\\
  &&\hspace{1em}
  \times\left[\frac{1}{(p_{1}-p_{2})^{2}-\mu^{2}}
    +\frac{1}{(l_{2}-l_{1})^{2}-\mu^{2}}\right]\nonumber\\
  &&\hspace{1em}\times \eta
  b_{i\lambda_{1}}^{\dag}(p_{1})
  d_{j\sigma_{1}}^{\dag}(l_{1})
  d_{j\sigma_{2}}(l_{2})
  b_{i\lambda_{2}}(p_{2})
  \eta\\
  \cG{2}^{\rm an}+\cG{2}^{{\rm an}\dag}
  &=&-g^{2}
  \sum_{\lambda_{1}}\sum_{\lambda_{2}}\sum_{\sigma_{1}}\sum_{\sigma_{2}}
  \int_{p_{1}}\int_{p_{2}}\int_{l_{1}}\int_{l_{2}}
  (2\pi)^{3}\delta^{3}(p_{1}+l_{1}-p_{2}-l_{2})\nonumber\\
  &&\hspace{1em}
  \times\bar{u}_{i}(p_{1},\lambda_{1})\gamma_{5}v_{i}(l_{1},\sigma_{1})
  \bar{v}_{i}(l_{2},\sigma_{2})\gamma_{5}u_{i}(p_{2},\lambda_{2})\nonumber\\
  &&\hspace{1em}
  \times\left[\frac{1}{(p_{1}+l_{1})^{2}-\mu^{2}}
    +\frac{1}{(p_{2}+l_{2})^{2}-\mu^{2}}\right]\nonumber\\
  &&\hspace{1em}\times \eta
  b_{i\lambda_{1}}^{\dag}(p_{1})
  d_{i\sigma_{1}}^{\dag}(l_{1})
  d_{i\sigma_{2}}(l_{2})
  b_{i\lambda_{2}}(p_{2})
  \eta,
\end{eqnarray}
which are all the second order interactions in the LF effective Hamiltonian.
Of course, since it does not have the particle-number-changing interactions,
the eigenstate is a pure two-body state.

We set the total transverse momentum $P_{\perp}$ to 0 for simplicity.
The eigenstate may be written as
\begin{eqnarray}
  |\Psi_{ij}(P^{+},m)\rangle&=&\sum_{\lambda_{1}}\sum_{\lambda_{2}}
  \int_{0}^{1}dx\int_{0}^{\infty}d\kappa\int_{0}^{2\pi}d\varphi
  \frac{1}{2(2\pi)^{3}}\sqrt{\frac{\kappa}{x(1-x)}}\nonumber\\
  &&\hspace{-3em}\times e^{i(m-\lambda_{1}/2-\lambda_{2}/2)\varphi}
  \Psi_{ij}(x,\kappa;\lambda_{1},\lambda_{2},m)
  b_{i\lambda_{1}}^{\dag}(p_{1})
  d_{j\lambda_{2}}^{\dag}(p_{2})
  |0\rangle,
\end{eqnarray}
where
\begin{equation}
  p_{1}^{+}=xP^{+},\quad p_{2}^{+}=(1-x)P^{+},\quad
  (p_{1\perp}-p_{2\perp})/2=\kappa_{\perp},
\end{equation}
\begin{equation}
  \kappa=|\kappa_{\perp}|,\quad
  \tan\varphi=\frac{\kappa_{\perp}^{2}}{\kappa_{\perp}^{1}}.
\end{equation}
$m$ is the eigenvalue of the third component of the total angular momentum 
$J_{3}$.
In the LF coordinates,
the squared total angular momentum \mbox{\boldmath $J$}$^{2}$
is not a good quantum number,
but the third component $J_{3}$, the helicity, is.

We discretized $x$ with $L$ equally-spaced points,
in the numerical calculations of diagonalization.
We put a transverse cutoff $\Lambda_{\perp}$ for $\kappa$ and,
also used the following variable
\begin{equation}
  z=\left(\frac{\kappa}{\Lambda_{\perp}}\right)^{1/3},
\end{equation}
instead of $\kappa$ and discretized $z$ with $N$ equally-spaced points
because the wavefunctions are sharp for $\kappa\sim 0$,
but flat for $\kappa\sim\Lambda_{\perp}$.

The results are shown in Fig.~\ref{fig:Result1} and Fig.~\ref{fig:Result2}.
We set the parameters $L=10$, $N=30$, $m=0$, the fermion mass $m_{1}=1.0$ GeV,
the anti-fermion mass $m_{2}=1.0$ GeV
and the scalar mass $\mu=0.01$ GeV in all the cases.
We define $\alpha_{g}=g^{2}/4\pi$.

Fig.~\ref{fig:Result1} shows $\alpha_{g}$ dependence of the binding
energy of the ground state for various transverse cutoff $\Lambda_{\perp}$.
(a) is the case that the fermion flavor is different from the anti-fermion one,
that is, excluding the fermion annihilation part.
(b) is the case of including it.
In (a), the binding energy is almost independent of $\Lambda_{\perp}$ in
$\alpha_{g}<1.5$, but the ground state is not a bound state
because it is slightly negative.
Even though the binding energy becomes positive and larger as
$\alpha_{g}$ grows, we can not say that the ground state is bound
because it apparently depends on $\Lambda_{\perp}$ for large $\alpha_{g}$.
This means that the binding energy depends on $\Lambda_{\perp}$
even if $\alpha_{g}$ is small.
If we consider the perturbation theory of the effective Hamiltonian,
which treats $\eta({\cal G}_{2}+{\cal G}_{2}^{\dag})\eta/2$ as the interaction,
with small $\alpha_{g}$,
the leading term of the eigenvalue which depends on the cutoff
corresponds to Eq.~(\ref{eqn:cF4-div})
times $(-1)$.
As mentioned in Sec.~\ref{sec:DivergenceHeff},
we expect that the dependence on $\Lambda_{\perp}$
is weakened in the next order calculation.
(b) has stronger dependence on $\Lambda_{\perp}$ than (a).

Fig.~\ref{fig:Result2} shows $\Lambda_{\perp}$ dependence of the binding
energy of the ground state for $g=3$ with the fermion annihilation part.
The reason for its behavior
is {\em not} that
the LF effective Hamiltonian needs the renormalization of the scalar mass
which corresponds to a fermion loop.
The perturbative analysis shows that the fourth order term of the eigenvalue
is logarithmically divergent like in
Eq.~(\ref{eqn:cF4-div}).
We consider that it is canceled by adding the fourth order terms.

\figput{fig:Result1}

\figput{fig:Result2}

\section{Ambiguity in the integration}\label{sec:Ambiguity}
In this appendix, we explain that 
there is in general an ambiguity in the integrations like 
that appearing in Eq.~(\ref{eqn:IncAmbiguity})
by demonstrating an explicit example.
Let us consider the following integration:
\begin{equation}
  I=\lim_{\Lambda_{1}\rightarrow\infty}
  \lim_{\Lambda_{2}\rightarrow\infty}
  \int_{-\Lambda_{1}}^{\Lambda_{1}}\frac{dq_{1}^{-}}{2\pi}
  \int_{-\Lambda_{2}}^{\Lambda_{2}}\frac{dq_{2}^{-}}{2\pi}
  \ed{k_{1}^{-}-q_{1}^{-}}
  q_{1}^{-}\frac{i}{q_{1}^{-}-q_{2}^{-}-i\epsilon}
  \ed{q_{2}^{-}-k_{2}^{-}}.
\end{equation}
If $k_{1}^{-}$ and $k_{2}^{-}$ are finite,
we can take the integration contours which have a infinite radius
in the lower half plane for $q_{1}^{-}$
and in the upper one for $q_{2}^{-}$ respectively.
By the residue theorem, $I$ becomes
\begin{equation}
  I=\frac{k_{1}^{-}+k_{2}^{-}}{2}\frac{i}{k_{1}^{-}-k_{2}^{-}-i\epsilon}
  -\lim_{\Lambda_{1}\rightarrow\infty}
  \lim_{\Lambda_{2}\rightarrow\infty}
  i\int_{0}^{-\pi}\frac{d\theta_{1}}{2\pi}
  \int_{0}^{\pi}\frac{d\theta_{2}}{2\pi}
  \frac{1}{1-\frac{\Lambda_{2}}{\Lambda_{1}}e^{i(\theta_{2}-\theta_{1})}},
\end{equation}
where we keep the radii finite in the second term
because $I$ have various values depending on ways to take the limits.
For example,
\begin{enumerate}

\item $\Lambda_{1}\rightarrow\infty$ before $\Lambda_{2}\rightarrow\infty$,
  \begin{equation}
    I=\frac{k_{1}^{-}+k_{2}^{-}}{2}\frac{i}{k_{1}^{-}-k_{2}^{-}-i\epsilon}
    +\frac{i}{4}.
  \end{equation}

\item $\Lambda_{2}\rightarrow\infty$ before $\Lambda_{1}\rightarrow\infty$,
  \begin{equation}
    I=\frac{k_{1}^{-}+k_{2}^{-}}{2}\frac{i}{k_{1}^{-}-k_{2}^{-}-i\epsilon}.
  \end{equation}

\item $\Lambda_{1}=\Lambda_{2}\rightarrow\infty$,
  \begin{equation}
    I=\frac{k_{1}^{-}+k_{2}^{-}}{2}\frac{i}{k_{1}^{-}-k_{2}^{-}-i\epsilon}
    +\frac{i}{8}.
  \end{equation}

\end{enumerate}
The other limits may yield the other constants. 
Note that such ambiguity arises from the terms whose total dimension
of the variables of the integrations is zero.

\end{document}